\documentclass[10pt]{article}
\usepackage{amssymb, amsthm, amsmath}

\usepackage{a4wide}
\usepackage{graphicx,pstricks}
\usepackage{pst-node,pst-3d,pst-text,pst-tree}
\usepackage{natbib}

\psset{linewidth=1pt,yunit=10mm,levelsep=10mm,tnsep=2pt}

\newcommand{\beqn}{\begin{eqnarray}\begin{aligned}}
\newcommand{\eqn}{\end{aligned}\end{eqnarray}}

\begin{document}

\begin{titlepage}

\begin{center}

{\Large {\bf Phylogenetic estimation with partial likelihood tensors}}

\vspace{2em}

J G Sumner$^{1,2}$ and M A Charleston$^{1,4,5}$
\par \vskip 1em \noindent
{\it $^1$School of Information Technologies, $^4$Centre for Mathematical Biology, $^5$Sydney Bioinformatics, University of Sydney, NSW 2006, Australia}\\
{\it $^2$School of Mathematics and Physics, University of Tasmania, TAS 7001, Australia}\\

\end{center}
%\footnotetext{ }
\par \vskip .3in \noindent

\vspace{1cm} \noindent\textbf{Abstract} \normalfont 
\\\noindent
We present an alternative method for calculating likelihoods in molecular phylogenetics.
Our method is based on partial likelihood tensors, which are generalizations of partial likelihood vectors, as used in Felsenstein's approach.
Exploiting a lexicographic sorting and partial likelihood tensors, it is possible to obtain significant computational savings.
We show this on a range of simulated data by enumerating all numerical calculations that are required by our method and the standard approach.

\vfill
\hrule \mbox{} \\
{\footnotesize 
{\textit{keywords:} phylogenetics, maximum likelihood, computational complexity, tensors}\\
{\textit{email:} jsumner@it.usyd.edu.au}\\
}
\end{titlepage}
\section{Introduction}
In his landmark paper \citep{felsenstein1981}, Felsenstein popularised the method of maximum likelihood for phylogenetic estimation.
Crucial to practical implementation was the introduction of a ``pruning algorithm'', which, under the assumptions of a reversible Markov process on a tree, allowed for efficient computation of the likelihood of observed molecular sequences.
Since this time, there has been an explosion in the use of the maximum likelihood method in phylogenetic studies. 
There have also been numerous algorithmic developments including computation with more general models \citep{boussau2006,yang1997} and heuristics speedups acting at the likelihood step \citep{guindon2003,stamatakis2006} or the tree search level \citep{guindon2003,whelan2007}. 
However, at the likelihood step, the basic algorithmic implementation still proceeds by applying Felsenstein's original recursive formula.

We present an alternative method for computing likelihoods given a tree, a root distribution and a set of transition matrices. 
The particulars of the transition matrices do not concern us, as our results are independent of the model of sequence evolution.
Hence, we will assume these matrices are given, and concentrate on computational complexity at the likelihood step.
This is well justified, as, aside from the tree search, the likelihood step is the most intensive part of maximum likelihood estimation \citep{bryant2005b}. 
Given the task of calculating of the likelihood of a single site in a sequence alignment, the method we present can actually be more cost intensive than applying Felsenstein's recursive formula. 
The effectiveness of our method becomes apparent by considering that in practice one is never calculating the likelihood of just a single site, but must calculate the likelihood of each and every site.
We will show that our method offers significant computational savings; a speedup of up to a factor of 6 for the most favourable of realistic cases.

As noted by Felsenstein, the most obvious cost-saving measure follows by observing that many site patterns in an alignment occur multiple times and there is no need to recalculate the likelihood each time.
The process of identifying these common sites is called ``aliasing''.
Aliasing aside, the basic premise of our method is that a large number of sites in an alignment are often very similar to each other (two sites share most states in common and differ on only 1 or 2 of the sequences). 
This is certainly true for realistic data sets, as they have evolved from a common ancestor in the not-too-distant past (at least by hypothesis if not in fact).
We define \emph{partial likelihood tensors} (PLTs) as multi-dimensional arrays that generalize Felsenstein's partial likelihood vectors (PLVs).
Using these PLTs, it becomes advantageous to sort the sites lexicographically and retain a few of these PLTs as the likelihood of each site is calculated. 
These PLTs can then be returned to when it comes to computing the likelihood of the next site, resulting in a minimization of the total number of calculations required. 
%This general approach, trading computations for memory, is known in the computer science literature as \emph{dynamic programming}.

An important aspect of our approach is that the lexicographic ordering can be computed exactly and efficiently using an $\mathcal{O}{\left(N(m+k)\right)}$ radix sort \citep{cormen2001}, where $N$ is the number of unique site patterns in the alignment, $m$ is the number of sequences and $k$ is the number of possible character states.
This should be compared to the related ``column sorting'' approach of \citet{pond2004}, where, as the calculation moves through the alignment, each PLV from the present site is retained.
Depending on how the next site differs from the present site, some of the PLVs required for the next site will be identical to those retained, and hence some superfluous computation can be avoided.
This approach relies upon a (heuristic) solution of a travelling salesman problem (TSP) to find an ordering of sites that maximizes the saving.
The solution of the TSP used by \citeauthor{pond2004} is $\mathcal{O}(N^2)$, so we can expect that their technique works best for shorter sequences.
This is the direct converse of the approach we present here, which is at its greatest power, relative to other approaches, when the sequence length is long, such that the data is very heterogeneous.

Another related approach is implemented by \citet{stamatakis2005a} using ``Subtree Equality Vectors'' (SEVs). 
This approach extends the idea of aliasing to the subtree level: 
faced with the need to calculate the likelihood on a given subtree, a sweep through the corresponding sequences in the alignment is performed, counting occurrences of \emph{homogeneous} subpatterns.
 (A homogeneous pattern is one in which each character state is identical.)
Only the homogeneous patterns are accounted for as a general count would not amortize well with large data sets, and the SEVs must be recomputed for each alternative tree.
\citeauthor{stamatakis2005a} are primarily interested in the problem of likelihood computations for data sets with very many taxa ($10^3$ and above).

% This section will introduce retroML using a septet tree, discuss how rank is a problem, give heuristic for choocing best starting leaf, give regML counts, show regML is indep of topology, give retroML counts, show that for a single site, retroML can take longer, show that for the caterpiller that retroML=regML and give the total cost of both thereof, introduce the concept of proportion of patterns.

\section{Methods}
Here we present our method, $\texttt{retroML}$, for computing the likelihood of molecular sequence data under the assumption of a Markov model on a rooted binary tree.
We do this with the aid of an example for a septet tree (Figure~\ref{fig:septet}). 
In Appendix~\ref{retroMLrestrict} we give an example for a tree with 16 leaves and in Appendix~\ref{pseudo} we give a generic presentation, valid for trees of any size.
We will also discuss the computational complexity of $\texttt{retroML}$ as judged against Felsenstein's approach $\mathcal{F}$.

\begin{figure}[tbp]
\centering
\includegraphics[scale=.5]{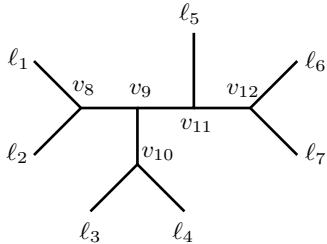} 
\caption{Septet tree}
\label{fig:septet}
\end{figure}

We consider an alignment of $m$ sequences, with no gaps. 
(It is straightforward to modify our results for when there are gaps, depending on how they are to be dealt with.)
A ``pattern'' will be the (ordered) sequence of states that occur at a given site in a sequence alignment.
The actual numeric values of the Markov model parameters will be of no concern to us: our results depend only on the number $m$ of leaves of the tree, its topology, the number $k$ of states, and the number $N$ of unique patterns in the alignment.
Thus, we take the transition matrices defining the Markov process on the tree as given (with one matrix for every vertex excluding the root) and consider the complexity of computing partial likelihood vectors at the root, conditioned on the patterns observed at the leaves.
Rather than present empirical timing results, which are dependent on computer hardware and/or programming language, we will compare an \emph{exact} count of numerical operations required of \texttt{retroML} to that of $\mathcal{F}$.
For this purpose, we define a ``cost'' of a computation as a pair representing the numbers of multiplications and additions required, respectively: $s(\cdot)\!=\![s^*(\cdot),s^+(\cdot)]$ for \texttt{retroML} and $f(\cdot)\!=\![f^*(\cdot),f^+(\cdot)]$ for the standard approach $\mathcal{F}$.

\subsection{Partial likelihood tensors}

In Felsenstein's method $\mathcal{F}$, a \emph{partial likelihood vector} (PLV) at a vertex represents the likelihood of observing each of the $k$ possible states at that vertex, conditional upon a pattern of states at the leaves of the subtree subtended by that vertex.
For the $i^\text{th}$ pattern, these PLVs are computed recursively by implementing the formula
\beqn
L_i^{v}(a)=\left( \sum_{b=1}^k M_{ab}^{u_1}L_i^{u_1}(b) \right) \cdot \left( \sum_{b'=1}^k M_{ab'}^{u_2}L_i^{u_2}(b') \right),\nonumber
\eqn
where $v$ is an internal vertex with children $u_1$ and $u_2$, and $M^{u_j}_{ab}$ is the probability of a transition from state $a$ to $b$ along the edge connecting $v$ and $u_j$.
For a PLV at a leaf, the entry that corresponds to the state present at that leaf is set to 1, whilst the other entries are set to 0.
This recursive formula plays a vital role in the efficient implementation of all likelihood calculations in molecular phylogenetics, and it is exactly this recursion that we aim to supplant with our approach.

The basic components underlying $\texttt{retroML}$ are \emph{partial likelihood tensors} (PLTs), which can be thought of as generalizations of partial likelihood vectors.
A PLT represents the likelihood of observing arbitrary states on \emph{multiple} vertices of a tree, conditioned upon observing certain states at some of the leaves.
Notationally, for a subset of vertices labelled as $(v_1,\ldots, v_r)$ with state $a_i$ at vertex $v_i$, and a subset of $\hat{m}\leq m$ leaves labelled as $(\ell_1,\ldots ,\ell_{\hat{m}})$ with state $X_i$ observed at leaf $\ell_i$, we express the corresponding PLT as
\beqn
\Psi_{a_1\ldots a_r}^{(v_1, \ldots ,v_r)}(X_1\ldots X_{\hat{m}}).\nonumber%:=\mathbb{P}[\chi(\ell_1)\!=\!X_1,\ldots ,\chi(\ell_{\hat{m}})\!=\!X_{\hat{m}};\chi(v_1)\!=\!a_1,\ldots \chi(v_r)\!=\!a_r],\nonumber
\eqn
The ``rank" of a PLT is here defined as the number of vertices $r$, and, given that the states $a_i$ are free to range over any of $k$ values, we see that a PLT encodes $k^r$ numbers.
Note that a PLV is a rank 1 PLT, but the converse is not true in general: a rank 1 PLT is not necessarily a PLV.

To judge performance, we will compare the cost of computing, at the root of a tree, the PLVs of the observed patterns in an alignment using $\mathcal{F}$ and $\texttt{retroML}$.
Equipped with these PLVs and a root distribution $\pi (a)$, the likelihood of each site is calculated the same way for both methods:
\beqn
L_i := \sum_{a=1}^k L_i^{(\texttt{root})}(a)\pi (a).\nonumber
\eqn
Finally, the negative $\log$-likelihood of the alignment is calculated by
\beqn
-\ln L := -\sum_{i=1}^N \alpha_i \ln L_i,\nonumber
\eqn
where $\alpha_i$ is the number of times the $i^{\text{th}}$ pattern occurs in the alignment.

Under conditions that we will clearly delineate in \S\ref{sec:simulations}, it is possible to achieve significant computational speedups by employing partial likelihood tensors.
This is made possible by observing that PLTs can be used to compute likelihoods that are conditioned upon \emph{subpatterns} of arbitrary length.
For example, in an alignment of 5 sequences with the pattern at the first site being $X_1X_2X_3X_4X_5$ it is possible that there will exist another pattern $X_1X_2X_3X_4'X_5'$ that has the same first three states.
%, elsewhere in the alignment, a pattern of the form $X_1X_2X_3\ast\ast$.
If we have the PLT that is conditioned on observing the subpattern $X_1X_2X_3$, then we see that we can save on computations by invoking this PLT when calculating the likelihood of $X_1X_2X_3X_4'X_5'$. 
This cannot be achieved with PLVs alone as the particular subpatterns they can incorporate are constrained by the tree topology.
An exception occurs when the tree topology is a ``caterpillar'' (completely unbalanced tree), for which $\texttt{retroML}$ can be implemented using PLVs only (we discuss this case in detail below).

We demonstrate $\texttt{retroML}$ with an example calculation on the septet tree in Figure~\ref{fig:septet}, retaining PLTs conditioned upon subpatterns of arbitrary length.
Without loss of generality, we consider $v_{12}$ to be the root of this tree and begin our computation at leaf $\ell_1$.
At each step we move to append the state at the next nearest leaf into the subpattern, as this helps to keep the rank of the PLTs minimal (this is an important consideration for reasons that we will discuss later).
Additionally, we retain only the PLTs that occur just before a leaf state is appended.
This ensures that when we return to a PLT for a subsequent pattern, the optimal saving in computation is achieved. 

\begin{figure}
\centering
\begin{tabular}{cc}
\includegraphics[scale=.5]{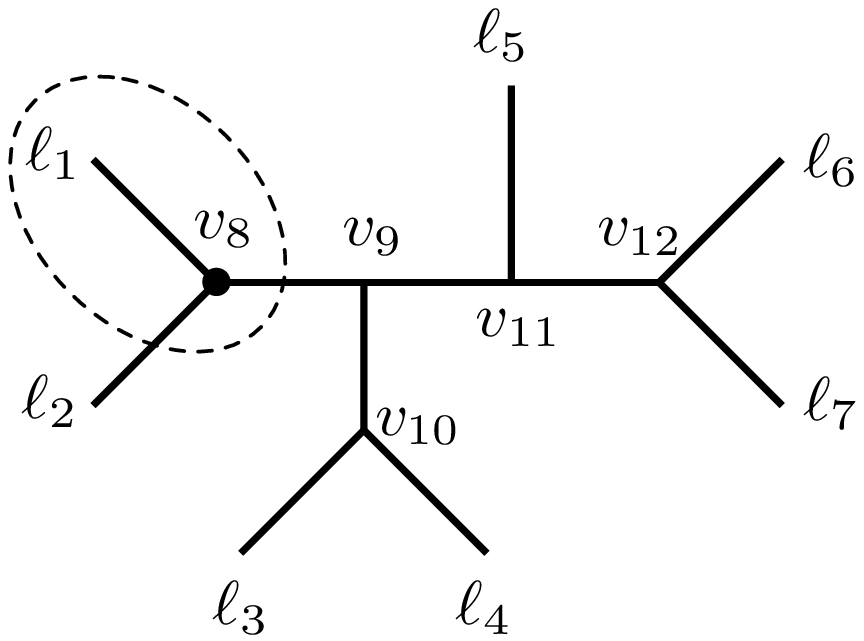} \hspace{5em} & \includegraphics[scale=.5]{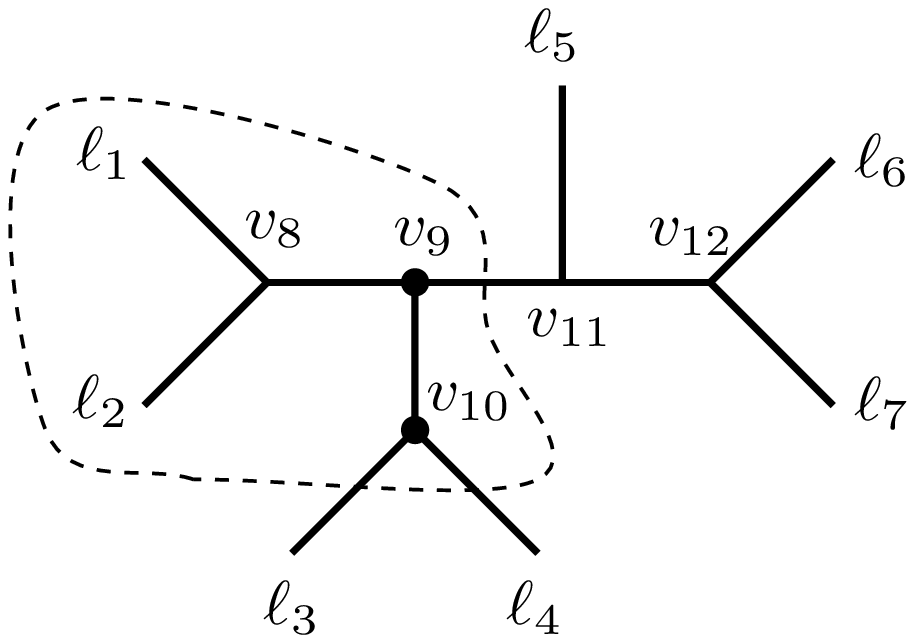} \\
\vspace{2em}(a) $\Psi^{(v_8)}_a(X_1)$\hspace{5em} & (b) $\Psi_{ab}^{(v_9,v_{10})}(X_1X_2)$ \\
\includegraphics[scale=.5]{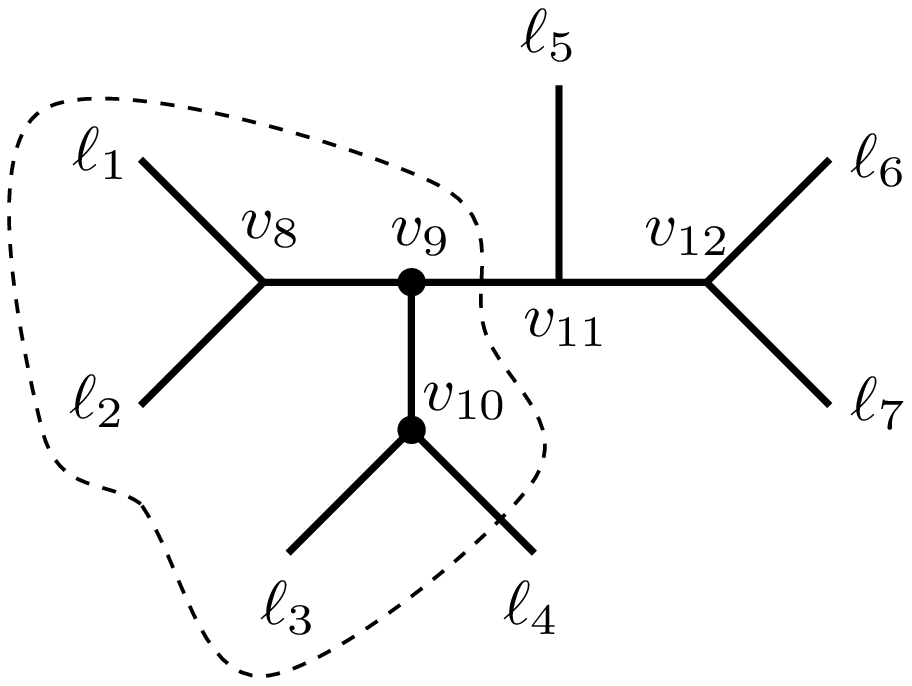}\hspace{5em} & \includegraphics[scale=.5]{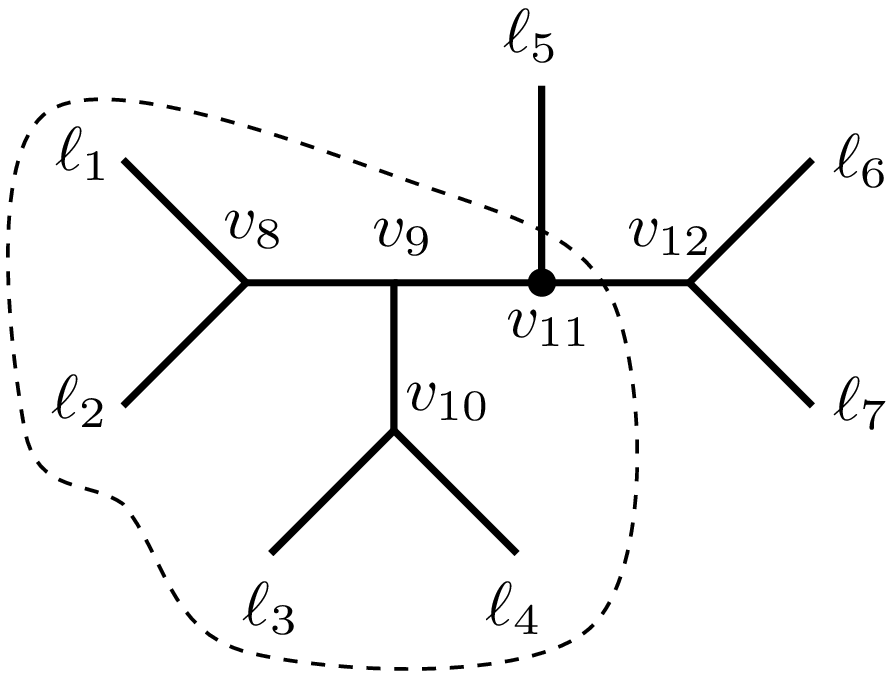}  \\
\vspace{2em}(c) $\Psi_{ab}^{(v_9,v_{10})}(X_1X_2X_3)$\hspace{5em} & (d) $\Psi_{a}^{(v_{11})}(X_1X_2X_3X_4)$ \\
\includegraphics[scale=.5]{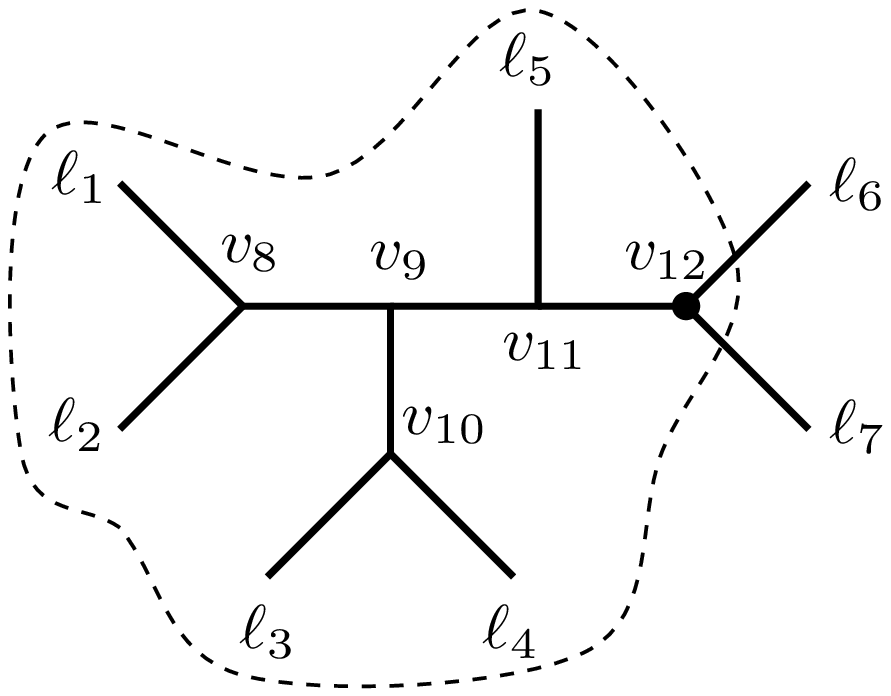}\hspace{5em} & \includegraphics[scale=.5]{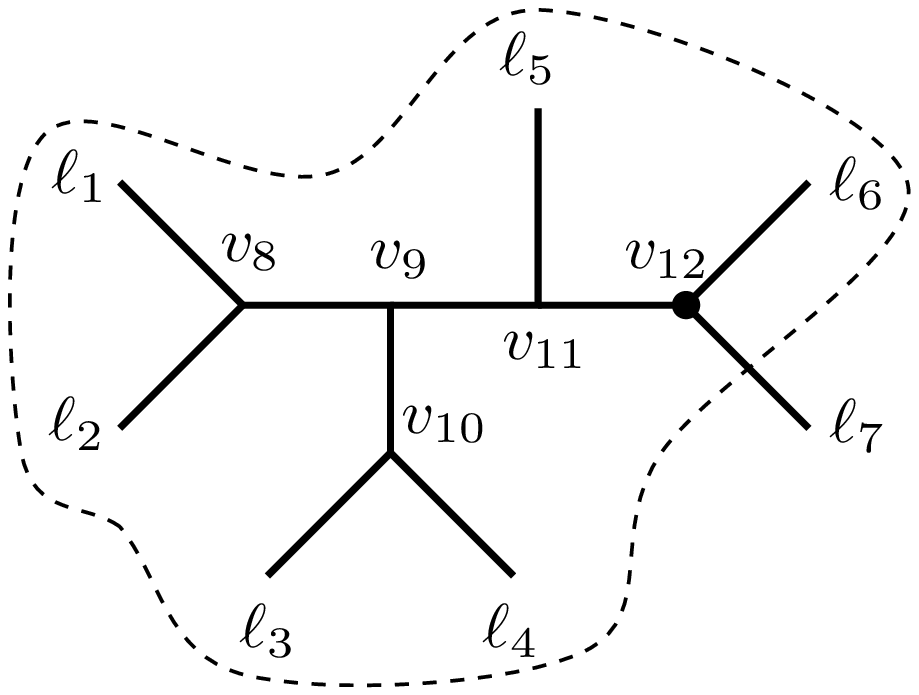}  \\
\vspace{2em}(d) $\Psi_{a}^{(v_{12})}(X_1X_2X_3X_4X_5)$\hspace{5em} & (e) $\Psi_{a}^{(v_{12})}(X_1X_2X_3X_4X_5X_6)$
\end{tabular}
\caption{Partial likelihood tensors}
\label{fig:PLTs}
\end{figure}

$\texttt{retroML}$ begins by computing the PLT conditioned on the observed state $X_1$ at $\ell_1$ and an arbitrary state at vertex $v_8$ (see Figure~\ref{fig:PLTs}(a)):
\beqn
\Psi^{(v_8)}_a(X_1)=M_{aX_1}^{v_1}.\nonumber
\eqn 
Note that this PLT is not a PLV, as it is conditional upon only the state at one of its children; the corresponding PLV at $v_8$ would be conditioned on the states at both children.
This PLT can be used to incorporate the subpattern $X_1X_2$ whilst moving over to $v_{10}$, retaining arbitrary states at $v_9$ and summing over the states at $v_8$ to give the rank 2 tensor (see Figure~\ref{fig:PLTs}(b))
\beqn
\Psi_{ab}^{(v_9,v_{10})}(X_1X_2)=\left( \sum_c M_{ac}^{v_8}\Psi^{(v_8)}_c(X_1)M_{cX_2}^{v_2} \right) M_{ab}^{v_{10}}.\nonumber
\eqn
From here, extending to the subpattern $X_1X_2X_3$ is a simple computation (see Figure~\ref{fig:PLTs}(c)):
\beqn
\Psi_{ab}^{(v_9,v_{10})}(X_1X_2X_3)=\Psi_{ab}^{(v_9,v_{10})}(X_1X_2)M_{bX_3}^{v_3}.\nonumber
\eqn
Here we see that the generality of the partial likelihood tensors is needed:
It is not possible to compute a likelihood for the subpattern $X_1X_2X_3$ using PLVs only; it is vital that arbitrary states at \emph{both} $v_9$ and $v_{10}$ are allowed for, and this requires a rank 2 tensor.
For this tree topology, there is simply no other way around this: without allowing for arbitrary states at $v_{10}$, one cannot incorporate the state $X_4$ into the final likelihood and, likewise, arbitrary states must be allowed at $v_9$ else the leaves $\ell_5,\ell_6$ and $\ell_7$ be neglected.
(These considerations are intimately tied to the \emph{separability} of the tensors in question \citep{landsberg2006,sumner2006a}.)

%\begin{figure}[tbp]
%\centering
%\includegraphics[scale=.5]{fig1a.eps} 
%\caption{$\Psi^{(v_8)}_a(X_1)$}
%\label{fig:septetA}
%\end{figure}
%
%\begin{figure}[tbp]
%\centering
%\includegraphics[scale=.5]{fig1b.eps} 
%\caption{$\Psi_{ab}^{(v_9,v_{10})}(X_1X_2)$}
%\label{fig:septetB}
%\end{figure}
%
%
%\begin{figure}[tbp]
%\centering
%\includegraphics[scale=.5]{fig1c.eps} 
%\caption{$\Psi_{ab}^{(v_9,v_{10})}(X_1X_2X_3)$}
%\label{fig:septetC}
%\end{figure}

Incorporating the subpattern $X_1X_2X_3X_4$, and then moving over to $v_{11}$ requires summing over the states at vertices $v_9$ and $v_{10}$ (see Figure~\ref{fig:PLTs}(d)):
\beqn
\Psi_{a}^{(v_{11})}(X_1X_2X_3X_4)=\sum_{a',b} M_{aa'}^{v_9}\Psi_{a'b}^{(v_9,v_{10})}(X_1X_2X_3)M_{bX_4}^{v_4}.\nonumber
\eqn
Note that this PLT is \emph{not} the PLV that would occur at $v_{11}$, as it is not conditioned upon $X_5$.

%\begin{figure}[tbp]
%\centering
%\includegraphics[scale=.5]{fig1d.eps} 
%\caption{$\Psi_{a}^{(v_{11})}(X_1X_2X_3X_4)$}
%\label{fig:septetD}
%\end{figure}

Extending to $X_1X_2X_3X_4X_5$ and moving over to $v_{12}$ requires summing over the states at $v_{11}$ (see Figure~\ref{fig:PLTs}(e)):
\beqn
\Psi_{a}^{(v_{12})}(X_1X_2X_3X_4X_5)=\sum_{a'}M_{aa'}^{v_{11}}\Psi_{a'}^{(v_{11})}(X_1X_2X_3X_4)M_{a'X_5}^{v_5}.\nonumber
\eqn
%Note that this PLT is exactly the PLV conditioned on $X_1$ to $X_5$ that the standard approach would compute at $v_{12}$.

%\begin{figure}[tbp]
%\centering
%\includegraphics[scale=.5]{fig1e.eps} 
%\caption{$\Psi_{a}^{(v_{12})}(X_1X_2X_3X_4X_5)$}
%\label{fig:septetE}
%\end{figure}

Incorporating the subpattern $X_1X_2X_3X_4X_5X_6$ is simple (see Figure~\ref{fig:PLTs}(f)):
\beqn
\Psi_{a}^{(v_{12})}(X_1X_2X_3X_4X_5X_6)=\Psi_{a}^{(v_{11})}(X_1X_2X_3X_4X_5)M_{aX_6}^{v_6},\nonumber
\eqn 
and, finally, we compute at $v_{12}$:
\beqn
\Psi_{a}^{(v_{12})}(X_1X_2X_3X_4X_5X_6X_7) = \Psi_{a}^{(v_{11})}(X_1X_2X_3X_4X_5X_6)M_{aX_7}^{v_7}.\nonumber
\eqn
One can check that this is exactly the PLV that Felsenstein's method $\mathcal{F}$ would calculate at $v_{12}$.

%\begin{figure}[tbp]
%\centering
%\includegraphics[scale=.5]{fig1f.eps} 
%\caption{$\Psi_{a}^{(v_{12})}(X_1X_2X_3X_4X_5)$}
%\label{fig:septetF}
%\end{figure}

From our procedure we see that we can save on computations if the above PLTs are retained for subsequent patterns.
For example, if the next pattern under consideration shares its first five states in common with a pattern that has already been dealt with, then we can return to $\Psi_{a}^{(v_{11})}(X_1X_2X_3X_4X_5)$ and the effective size of the tree that must be traversed to calculate the PLV of the pattern is only 2.

This concept will hold in general for $\texttt{retroML}$: 
%\begin{center}
%\fbox{\parbox[t]{40em}{
If the pattern under consideration shares its first $\hat{m}$ states in common with a pattern for which the likelihood has already been computed, the effective size of the tree that must be traversed is $(m\!-\!\hat{m})$.
%}}
%\end{center}
Clearly, this approach has the potential to save on significant amounts of computation.

As we discuss in detail below, the attractiveness of the approach is offset by the number of computations required in each step, which is $\mathcal{O}(k^r)$, where $r$ is the rank of the PLT involved, and, for worst-case tree topologies, this rank can become prohibitively large.
This is exactly what Felsenstein's post-order recursive method $\mathcal{F}$ avoids.
A secondary issue for $\texttt{retroML}$ is that the memory requirements of keeping all these PLTs in memory could easily become prohibitive for a large number of observed patterns.
This issue is easily addressed: by tackling the observed patterns in a certain order, only the $(m\!-\!1)$ PLTs from the \emph{previous} pattern need to be retained in memory.
We next describe how this can be achieved.

\subsection{Lookbacks}\label{sec:lookbacks}

We refer to an ordering of the leaves of a tree as a ``leaf order''.
For a tree with $m$ leaves, there are of course $m!$ possible leaf orders.
The leaf order will be set by the order in which our algorithm visits the leaves, and thus sets the order of the states in the patterns.

Given a list of patterns and a leaf order, a ``lookback value'' $\sigma_i$ is $m$ minus the position of the first state (counting from 0) where the $(i\!-\!1)^\text{th}$ and $i^\text{th}$ patterns differ, with $\sigma_1\!=\!m$. 
The lookback value of the current pattern tells us exactly which PLT to return to: there is no additional computation involved.
Additionally, considered as character strings, a lexicographic ordering of patterns will minimize the sum of the lookback values. 
This follows from the definition of lexicographic ordering. 
Since the effective tree size is reduced to $\sigma_i$, our approach will perform very well when the pattern under consideration has a small lookback value.
If the patterns are sorted lexicographically, we see that only the PLTs from the previous pattern need be retained in memory for best-performance to be attained.
This also solves the secondary issue raised above, as retaining only the PLTs from the previous pattern in memory keeps the memory requirements constant.
 
To get an idea of the distribution of lookback values that may occur in an alignment, we consider the idealised case where \emph{every} possible pattern occurs.
For an alignment of $m$ sequences, the number of possible patterns is $k^m$.
If the patterns are sorted lexicographically, then the number of patterns with lookback value $\sigma\!=\!m$ is $k$ and the number with lookback value $0< \sigma \leq (m\!-\!1)$ is $k^{m\!-\!\sigma}(k\!-\!1)$.
We note that 
\beqn
k+\sum_{\sigma=1}^{m-1}k^{m\!-\!\sigma}(k\!-\!1)=k+(k^m\!-\!k^{m\!-\!1})+(k^{m\!-\!1}\!-\!k^{m\!-\!2})+\ldots +(k^3\!-\!k^2)+(k^2\!-\!k)=k^m.\nonumber
\eqn

In the above example we took a particular traversal of the tree, allowing for the computation of the PLTs conditioned on subpatterns of sizes 1 to 7. 
The performance of $\texttt{retroML}$ is highly dependent on the way in which the tree is traversed; at all times it is important to minimize the rank of the resulting PLTs.
For a given tree, there is an optimal path that keeps the rank of the PLTs minimal and, in general, this path is not unique. 
For instance, in the above example, there would be no difference in the ranks attained if we started at any of the leaves $\ell_1,\ell_2,\ell_3,\ell_4,\ell_6,\ell_7$: in these cases the highest rank attained is 2 and this occurs only once.
If, however, we started at $\ell_5$ a rank 2 PLT would be required \emph{twice}: once on the vertices $v_9,v_{10}$ and once on the vertices $v_{11},v_{12}$.
Additionally, the performance of $\texttt{retroML}$ depends on the structure of the observed patterns after sorting; the smaller the value of the lookbacks, the better.
We showed above that if all the possible patterns are present, then the number of patterns with a lookback of $\sigma$ is $\mathcal{O}(k^{m\!-\!\sigma\!+\! 1})$, from which we see that if high rank PLTs occur in the implementation then it is better that they are encountered \emph{early}, for large lookback values.
For instance in our example, starting at any of the leaves $\ell_1,\ell_2,\ell_3,\ell_4$ will outperform starting at $\ell_6$ or $\ell_7$, because in the former the rank 2 PLT is encountered when $\sigma\!=\! 5$, whereas in the later the rank 2 PLT is encountered when $\sigma\!=\! 4$.

In any practical implementation, the interaction between the dependence on tree topology and the lookback structure of patterns is complicated.
We will show in \S\ref{sec:simulations} that the performance for different tree topologies is quite distinct; the ranks of the PLTs attained tell all.
Whereas, the effect of the observed patterns is dominated by their number; judged only by their lookback values, the patterns occurring in sequence alignments are more-or-less random.

We have found a particular tree traversal that takes into account these issues and works well in practice.
We do not prove that this is the best possible tree traversal, but provide it in Appendix~\ref{pseudo} as a useful heuristic.
It is designed to ensure that the rank of the PLTs encountered is kept minimal by at all times ensuring the method moves to the next nearest leaf.
In Appendix~\ref{pseudo} we also supply a useful heuristic for finding the best starting leaf based on the above considerations.

\subsection{Costs}
In this section we detail the exact costs of each of the steps involved in $\mathcal{F}$ and $\texttt{retroML}$.
Because the particular model parameters present are of no consequence to the cost of computations, in this section we will omit all vertex labels from transition matrices, PLVs and PLTs. 
We also omit the site label $i$ from the PLVs.

\begin{figure}[tbp]
\centering
\begin{tabular}{ccc}
\includegraphics[scale=.5,viewport=30 -30 120 90]{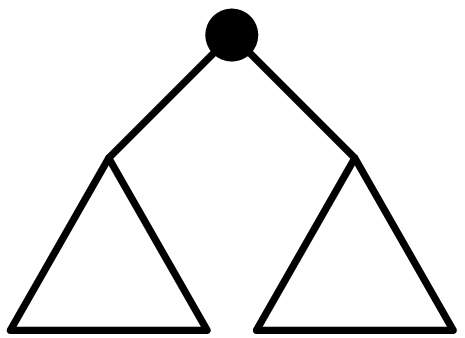} \hspace{3em} &  \includegraphics[scale=.5]{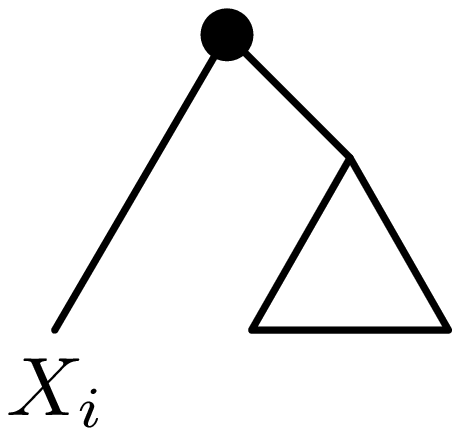} \hspace{3em} & \includegraphics[scale=.5]{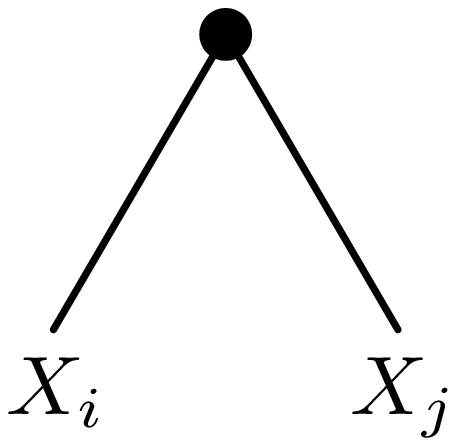}  \\
(a) \hspace{3em} & (b)  \hspace{3em} & (c) 
\end{tabular}
\caption{(a) An ``internal'' vertex; (b) A ``half internal'' vertex; (c) A ``cherry'' vertex}
\label{fig:vertextype}
\end{figure}

So that we are fair in the comparisons we make, we note here that the computation of a partial likelihood vector at a vertex can be simplified if one or both of its children are leaves.
In the case that the vertex is completely internal (i.e.\ neither child is a leaf; see Figure~\ref{fig:vertextype}(a)) the calculation proceeds as 
\beqn
L^{\texttt{int}}(a)=\left( \sum_{b} M_{ab}L(b) \right) \cdot \left( \sum_{c} M_{ac}L(c) \right),\nonumber
\eqn
which, to compute for $1\leq a\leq k$, costs 
\[
f(L^{\texttt{int}})=[k(2k\!+\! 1),2k(k\!-\!1)].
\]
However, if exactly one of the children of the vertex is a leaf (see Figure~\ref{fig:vertextype}(b)), then the calculation is
\beqn
L^{\texttt{halfint}}(a)=\left( \sum_{b} M_{ab}L(b) \right) \cdot \left( \sum_{c} M_{ac}\delta_{cX} \right),\nonumber
\eqn
which can be simplified to
\beqn
L^{\texttt{halfint}}(a)=\left( \sum_{b} M_{ab}L(b) \right) \cdot M_{aX},\nonumber
\eqn
with cost 
\[
f(L^{\texttt{halfint}})=[k(k\!+\! 1),k(k\!-\!1)].
\]
Similarly, if both children are leaves (making a ``cherry''; see Figure~\ref{fig:vertextype}(c)), the calculation simplifies further:
\[
L^{\texttt{cherry}}(a)=M_{aX_1}M_{aX_2}
\]
with cost 
\[
f(L^{\texttt{cherry}})=[k,0].
\]
Finally, if we place the root of a tree at a trifurcating vertex, we find in a similar manner that, if the root has $w\!=\!1$ or 2 children that are leaves, the cost is
\beqn
f(L^{\texttt{root}})=\left[k((3\!-\!w)k\!+\! 2),k(3\!-\!w)(k\!-\!1)\right].\nonumber
\eqn

%\begin{figure}[thp]
%\centering
%\includegraphics[scale=.5]{fig2a.eps} 
%\caption{An ``internal'' vertex}
%\label{fig:internal}
%\end{figure}
%
%\begin{figure}[thp]
%\centering
%\includegraphics[scale=.5]{fig2b.eps} 
%\caption{A ``half internal'' vertex}
%\label{fig:halfinternal}
%\end{figure}
%
%\begin{figure}[thp]
%\centering
%\includegraphics[scale=.5]{fig2c.eps} 
%\caption{A ``cherry'' vertex}
%\label{fig:cherry}
%\end{figure}

From these costs we see that the total cost of $\mathcal{F}$ for single site on a caterpillar tree (one cherry, $(m\!-\!4)$ half-internal vertices and a trifurcating root with 2 leaves) is 
\beqn
f(\texttt{singlesite})=&\left[k\!+\! (m\!-\!4)k(k\!+\! 1)\!+\! k(k\!+\! 2) , 0+(m\!-\!4)k(k\!-\!1)+k(k\!-\!1) \right]\nonumber\\
=&\left[(m\!-\!1)k\!+\! (m\!-\!3)k^2,(m\!-\!3)k(k\!-\!1)\right].
\eqn
This result is actually the same for all tree topologies.
As any tree topology can be changed to any another by repeatedly removing single leaves and re-inserting them elsewhere on the tree, we show presently that this does not affect the total cost $f(\texttt{singlesite})$.
In the case where a leaf is removed from a cherry vertex whose parent is an internal vertex, the difference in cost is 
\beqn
f(L^{\texttt{cherry}})+f(L^{\texttt{int}})-f(L^{\texttt{halfint}})=\left[ k(k\!+\! 1),k(k\!-\!1)\right].\nonumber
\eqn
If the leaf is removed from a triplet--from the cherry or otherwise, then the difference in cost is
\beqn
f(L^{\texttt{halfint}})+f(L^{\texttt{cherry}})-f(L^{\texttt{cherry}})=\left[ k(k\!+\! 1),k(k\!-\!1)\right].\nonumber
\eqn
So, no matter where the leaf is removed from, the difference in cost is the same.
Now, by observing that the insertion of a leaf is exactly equivalent to the reverse of one these cases, we see that the cost of $\mathcal{F}$ is independent of tree topology.
(See Figure \ref{fig:vertextype} for an illustration of this.)

Incidentally, if these simplifications are not taken into account, each PLV calculation takes the time of the PLV at a fully internal vertex.
If this were the case, then (asymptotically in $m$) the calculation of the likelihood of each pattern would be slowed by a factor of  $(2k\!+\! 1)/(k\!+\! 1)$, which for $k\!=\!4$ is 1.8.
We do not know if the standard phylogenetics maximum likelihood software packages take account of these simplifications or not.
It is clear that \citet{pond2004} did not take account of this in their analysis: they assigned a unit cost to each ``tainted'' vertex regardless of whether the vertex was internal, half internal or above a cherry.
Clearly, taking these factors into account would affect the solution of their TSP problem and consequently the optimal column sorting for a given data set.

The situation for our algorithm $\texttt{retroML}$ is quite different as the cost is \emph{not} independent of the tree topology; the extra information that is contained in the partial likelihood tensors allowing the lookback step comes at additional computational cost.
This additional cost depends on the rank of the tensor.

\begin{figure}[tbp]
\centering
\begin{tabular}{cc}
\hspace{-3em} 	\includegraphics[scale=.5]{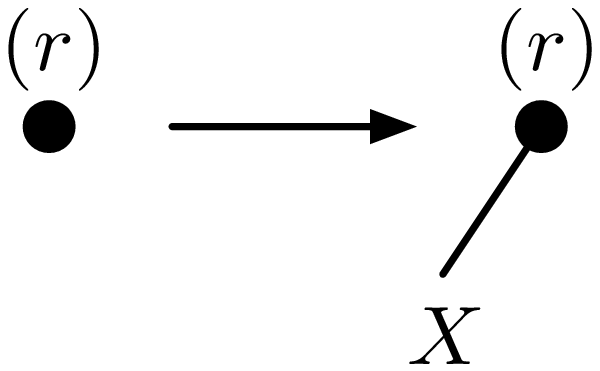}  &  \hspace{3em} \includegraphics[scale=.5,viewport=30 -10 120 120]{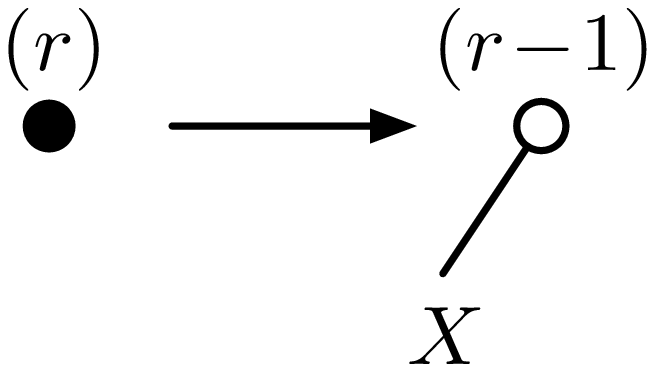}  \\
\hspace{-3em} (a) & \hspace{3em}  (b) \vspace{1Em}\\
\hspace{-3em} \includegraphics[scale=.5]{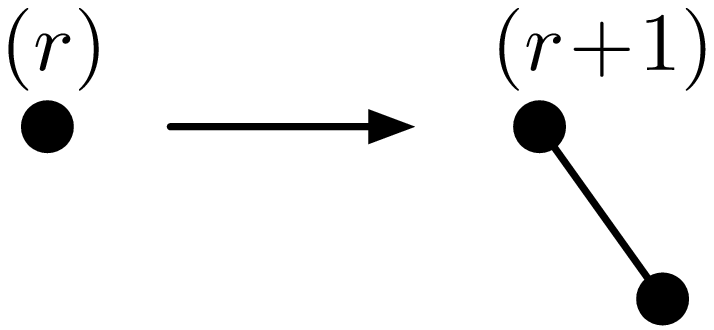} & \hspace{3em}  \includegraphics[scale=.5,viewport=30 -10 120 120]{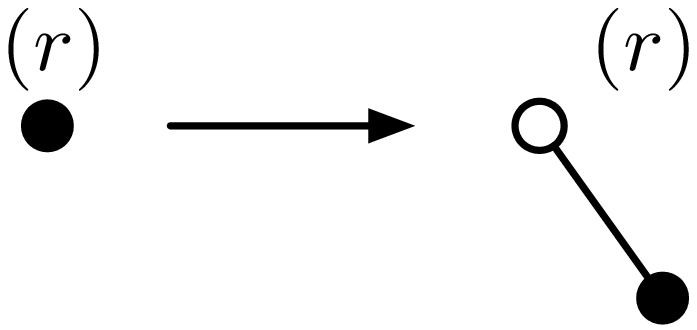}  \\
\hspace{-3em} (c) & \hspace{3em}  (d) 
\end{tabular}
\caption{(a) Evaluate leaf state whilst retaining vertex state; (b) Evaluate leaf state and sum over vertex states; (c) Move to new vertex whilst retaining state at current vertex; (d) Sum over states at vertex and move to adjacent vertex.}
\label{fig:moves}
\end{figure}

If we start the algorithm at a leaf vertex, the first step costs nothing:
\beqn
\Psi_a(X_1)=M_{aX_1};\nonumber
\eqn
it is simply an assignment.
After this beginning we assume that the current PLT is rank $r$ and there are four available generic moves:
(a) and (b) apply when the current vertex is directly above a leaf, and (c) and (d) apply when the current vertex is completely internal.
\begin{enumerate}
\item[(a)] Evaluate observed state at a leaf whilst retaining state at current vertex (see Figure~\ref{fig:moves}(a)): 
\[
\Psi_{a_1a_2\ldots a_r}\leftarrow \Psi_{a_1a_2\ldots a_r} M_{a_1X}.
\] 
This costs $s(\Psi_{a_1a_2\ldots a_r})\!=\![k^r,0]$ and the rank is unchanged: $r \leftarrow r$.
\item[(b)] Evaluate observed state at a leaf and sum over state at current vertex (see Figure~\ref{fig:moves}(b)): 
\[
\Psi_{a_2\ldots a_r}\leftarrow \sum_{a_1}\Psi_{a_1\ldots a_r}M_{a_1X}.
\]
This costs $s(\Psi_{a_2\ldots a_r})\!=\![k^r,k^{r\!-\!1}(k\!-\!1)]$ and the rank is reduced by one: $r \leftarrow r\!-\!1$.
\item[(c)] Move to adjacent vertex whilst retaining state at current vertex (see Figure~\ref{fig:moves}(c)): 
\[
\Psi_{aa_1a_2\ldots a_r} \leftarrow \Psi_{a_1a_2\ldots a_r} M_{a_1a}.
\]
This costs $s(\Psi_{aa_1a_2\ldots a_r})\!=\![k^{r\!+\! 1},0]$ and the rank is increased by one: $r \leftarrow r\!+\! 1$.
\item[(d)] Sum over states at vertex and move to adjacent vertex (see Figure~\ref{fig:moves}(d)):  
\[
\Psi_{aa_2\ldots a_r} \leftarrow \sum_{a_1}\Psi_{a_1a_2\ldots a_r} M_{a_1a}.
\]
This costs $s(\Psi_{aa_2\ldots a_r})\!=\![k^{r\!+\! 1},k^r(k\!-\!1)]$ and the rank is unchanged: $r \leftarrow r$.
\end{enumerate}

%\begin{figure}[tbp]
%\centering
%\includegraphics[scale=.5]{fig3a.eps} 
%\caption{Evaluate leaf state whilst retaining vertex state}
%\label{fig:leafretain}
%\end{figure}
%
%\begin{figure}[tbp]
%\centering
%\includegraphics[scale=.5]{fig3b.eps} 
%\caption{Sum over state at vertex and move to adjacent vertex}
%\label{fig:vertexsum}
%\end{figure}
%
%\begin{figure}[tbp]
%\centering
%\includegraphics[scale=.5]{fig3c.eps} 
%\caption{Evaluate leaf state and sum over vertex states}
%\label{fig:leafsum}
%\end{figure}
%
%\begin{figure}[tbp]
%\centering
%\includegraphics[scale=.5]{fig3d.eps} 
%\caption{Move to new vertex whilst retaining state at current vertex}
%\label{fig:vertexretain}
%\end{figure}

We note that in each of these moves for $\texttt{retroML}$, as well as for the steps of $\mathcal{F}$, the number $s^+,f^+$ of additions is strictly less than the number $s^\ast,f^\ast$ of multiplications.
Thus, for ease of presentation, from here on we will keep track of (and compare) counts of multiplications only.

Finally, we note that the lexicographic ordering of patterns can be achieved using a radix sort \citep{cormen2001}, which is $\mathcal{O}(m(N\!+\! k))$.
We assume that the time it takes to do this sort will not be significant to our analysis.
We justify this by noting that, for each candidate tree, the lexicographic ordering of the data set need only be computed once, whereas to optimize model parameters the likelihood calculation must be iterated many times. 
In fact, the number of iterations that must be performed on a given tree scales with the number of edges on the tree.
Thus, accounting for these iterations, the complexity of the standard implementation $\mathcal{F}$ is $\mathcal{O}(Nm^2k^2)$.

We can also counterpoint the cost of ordering the data set to the ``column sorting'' performed by \citet{pond2004}.
In that approach, an $\mathcal{O}(N^2)$ \emph{approximate} solution to a TSP was required to sort the data set.
Given that in most applications $m<N$, we can argue strongly that if the column sorting approach can achieve significant speedups, regardless of the need to solve the TSP, we certainly expect that our approach will do so also.

Breaking tack from the main thread of this article, which favours exact counts of empirical timings, we tested these assertions by timing likelihood and pattern sorting computations on a personal computer.
For instance, while the likelihood calculation using PAUP$^\ast$ \citep{paup} on an arbitrary tree constructed from a plant data set \citep{goremykin2003} of 15 sequences and 31k base pairs took approximately 0.33s, sorting the patterns in R \citep{Rproject} took approximately 0.005s on the same machine.
This was without taking advantage of the possibilities of radix sorting.

\section{Results}

\subsection{Simulation study}\label{sec:simulations}

Using the counts presented in the previous section, we conducted a simulation study comparing the number of multiplications required by $\texttt{retroML}$ to that of Felsenstein's method $\mathcal{F}$.
We generated DNA alignments with Filo \citep{filo} for each tree size in the range $m\!=\!4$ to 15 and each sequence length in the range $1$ to $10^4$.
The trees were randomly generated using the Yule model (birth only process with pendant edge lengths drawn from a uniform distribution).
We used the HKY Markov model with flat parameter settings and imposed a molecular clock with root-to-leaf height of .32 substitutions per site.
For each sequence alignment we found the multiplicative cost of computing likelihoods at the root on \emph{another} randomly tree generated under the same conditions.
The reason we generated a new tree is that we did not want there to be any biasing effect from the lookback structure caused by the true tree (in unpublished tests we showed that any such effect is not detectable anyway.)

\begin{figure}[tbp]
\centering
\begin{tabular}{cc}
\includegraphics[height=0.28\textheight,width=0.32\textheight]{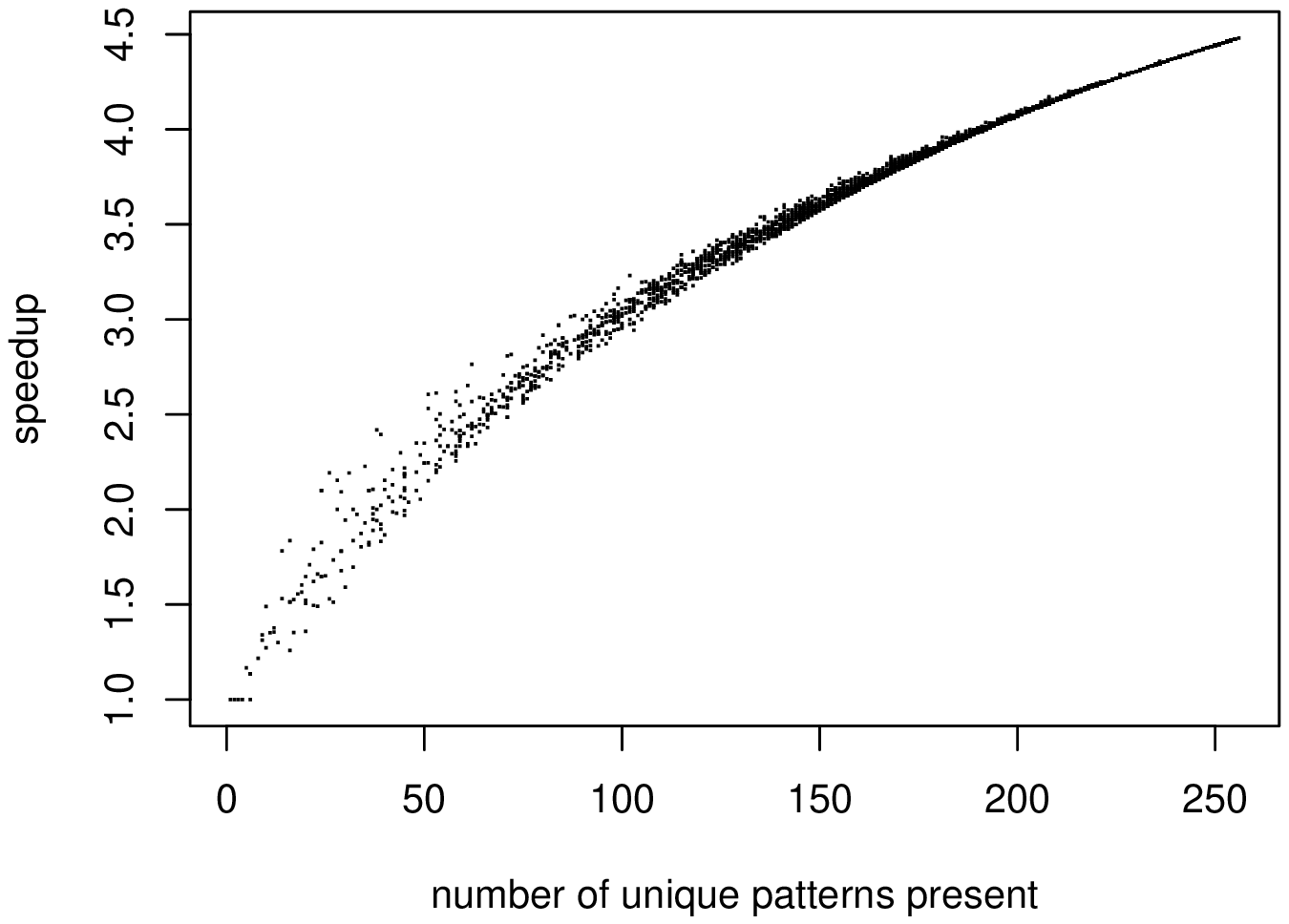}  & \includegraphics[height=0.28\textheight,width=0.32\textheight]{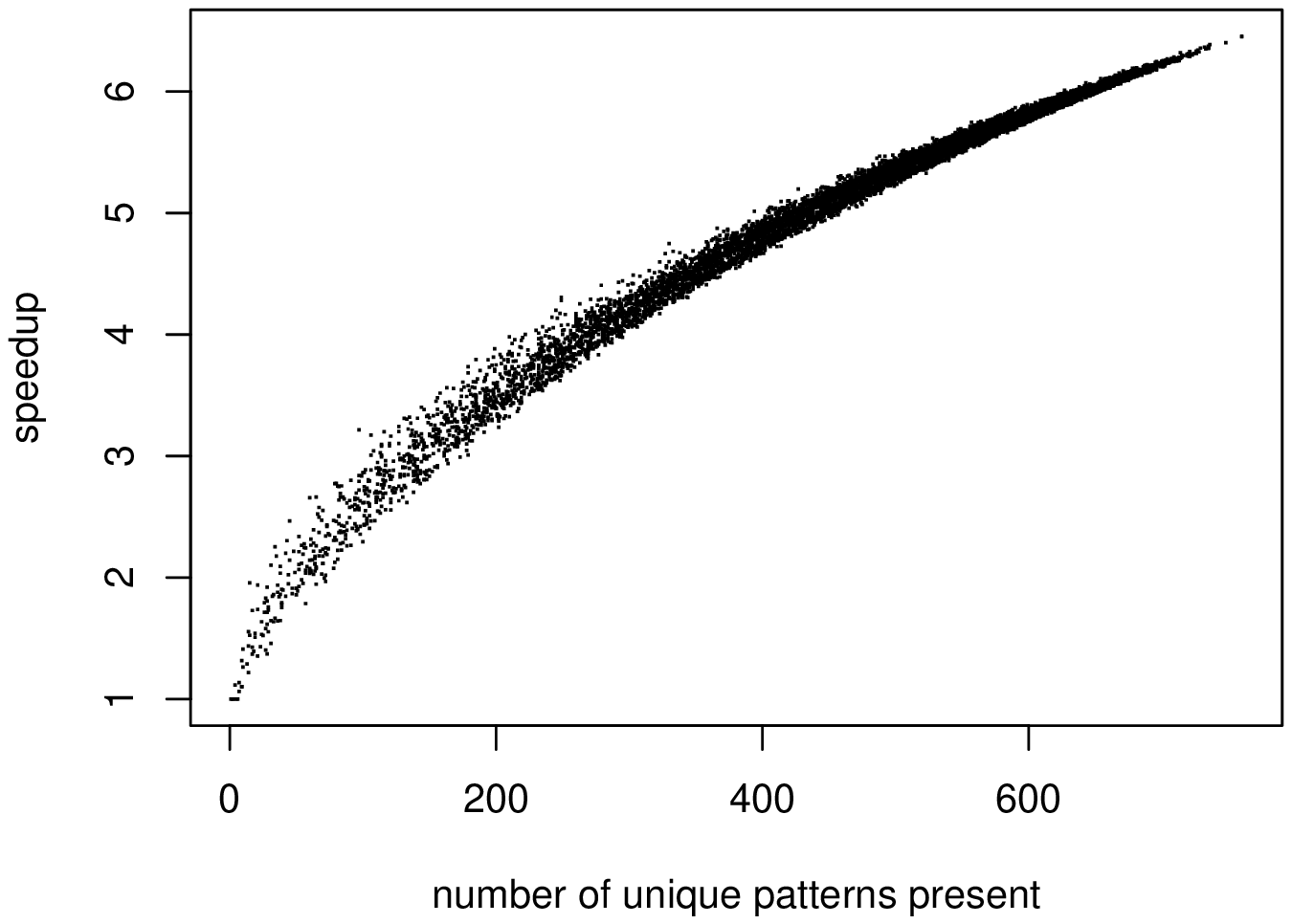} \\
(a) quartets, $m\!=4\!$ & (b) quintets, $m\!=5\!$ \\
\includegraphics[height=0.28\textheight,width=0.32\textheight]{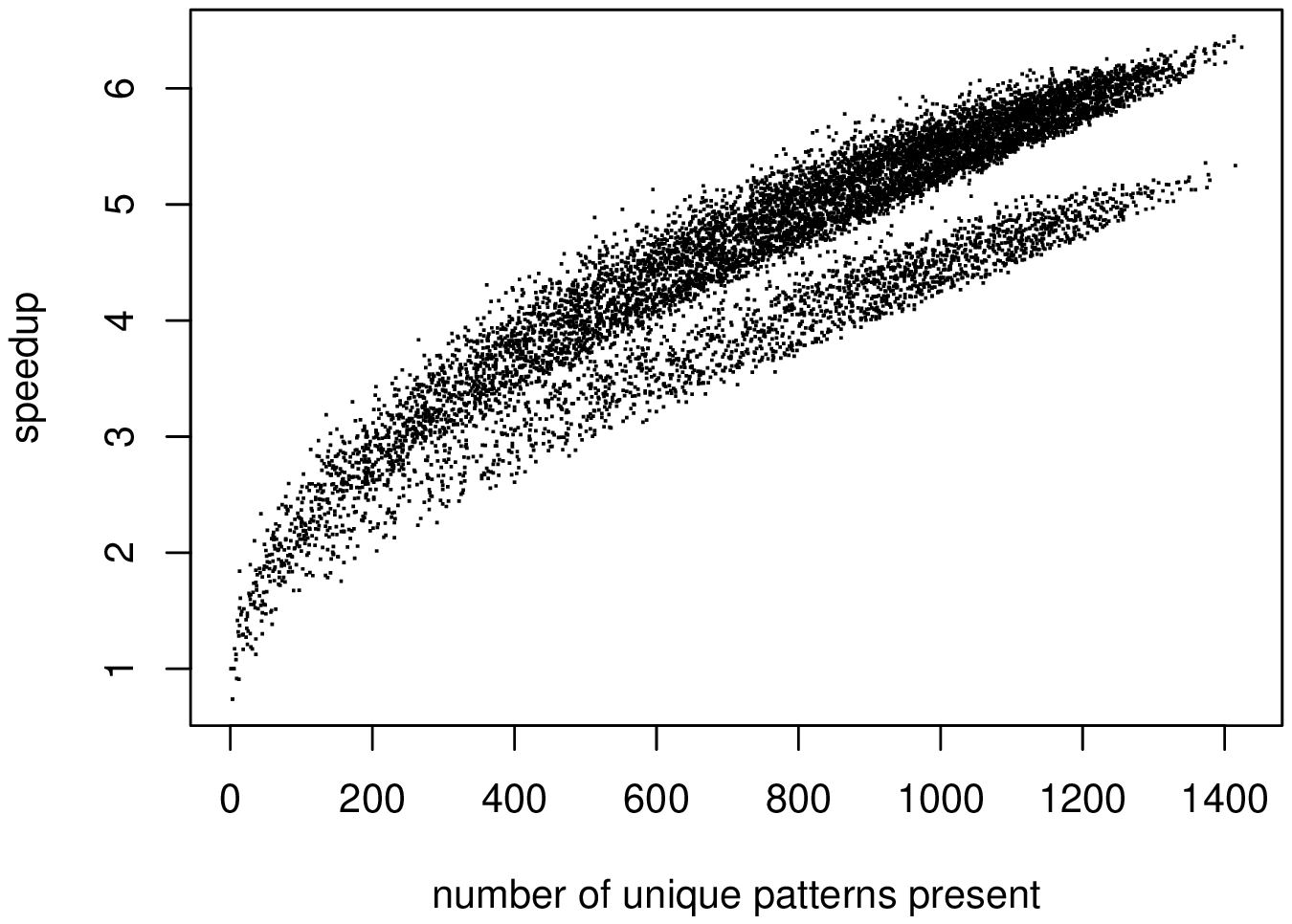}  & \includegraphics[height=0.28\textheight,width=0.32\textheight]{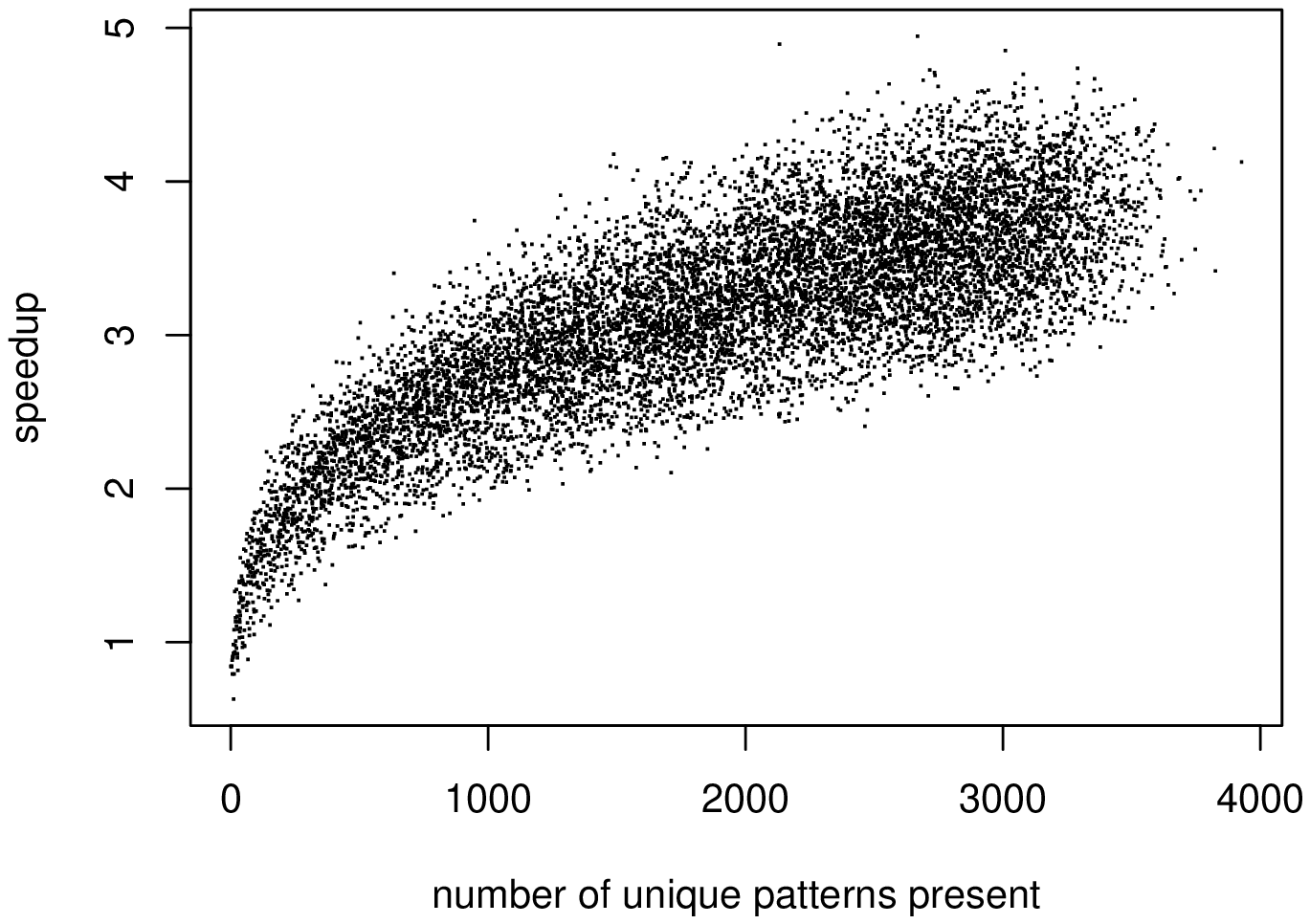} \\
(c) sextets, $m\!=6\!$ & (d) nonets, $m\!=9\!$ \\
\includegraphics[height=0.28\textheight,width=0.32\textheight]{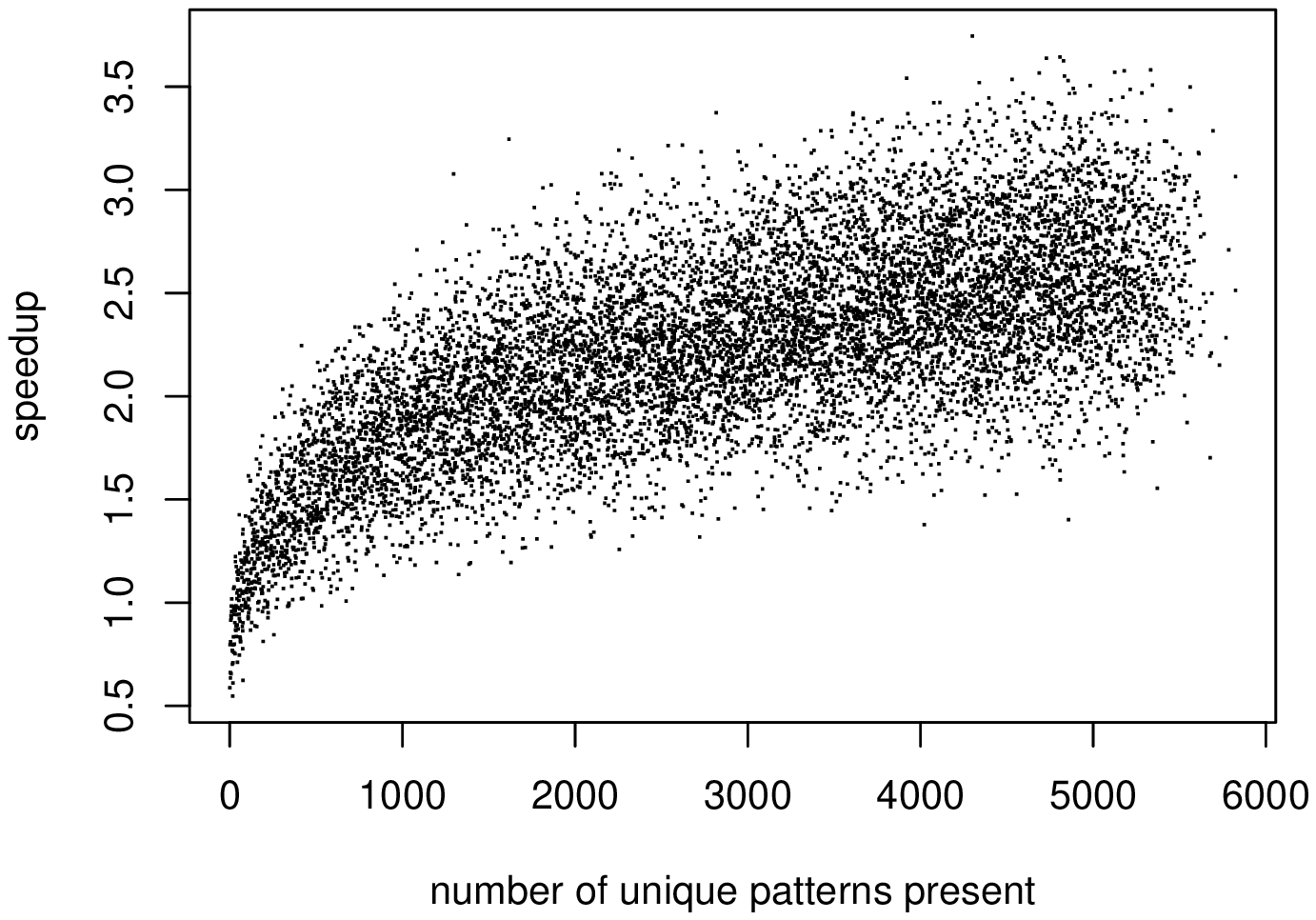}  & \includegraphics[height=0.28\textheight,width=0.32\textheight]{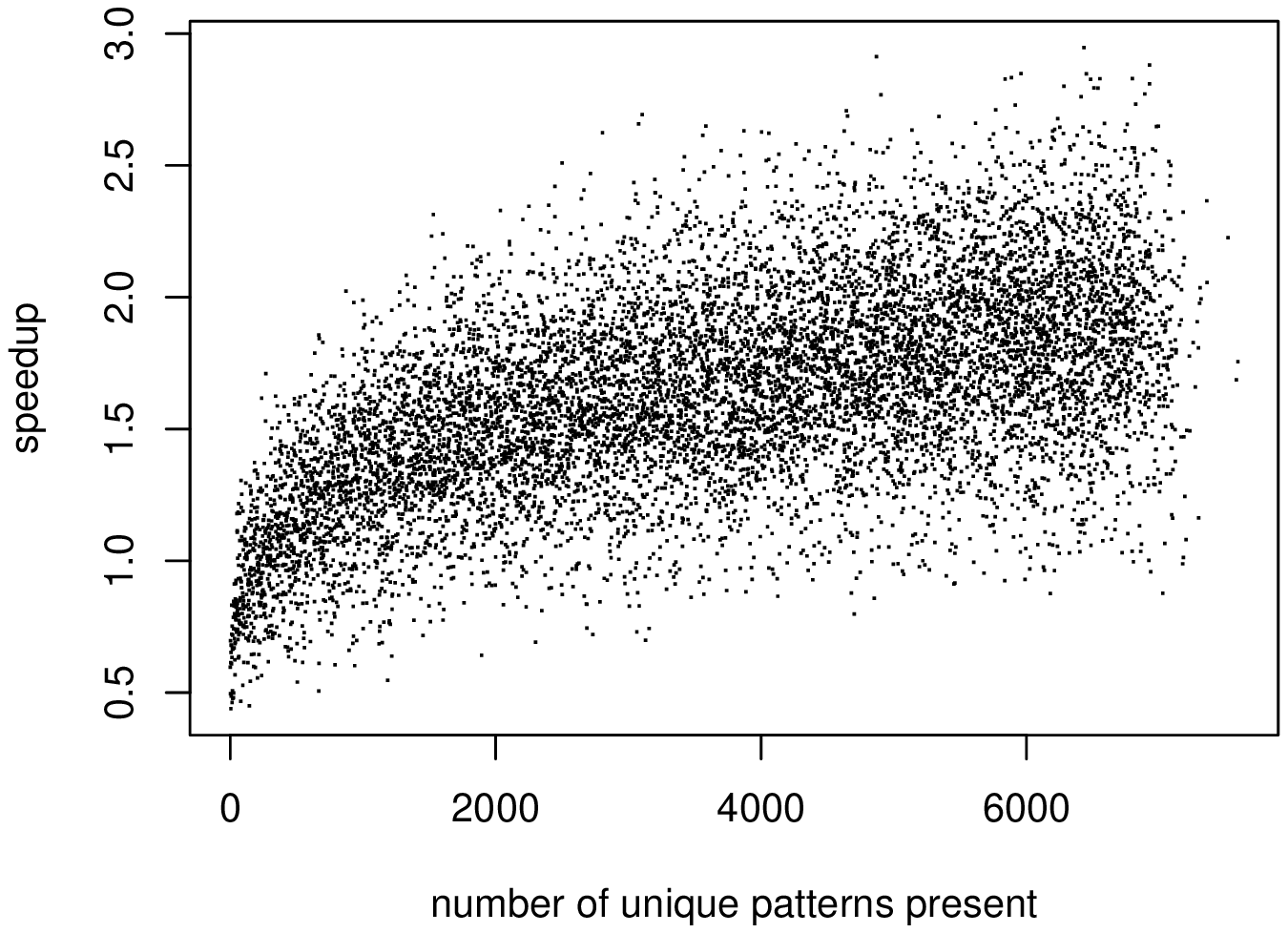} \\
(e) $m\!=12\!$ & (f)  $m\!=15\!$ \\
\end{tabular}
\vspace{1em}
\caption{Speedups ${f^\ast}/{s^\ast}$ for sequence length 1 to $10^4$.}
\label{fig:simulations}
\end{figure}

In Figure~\ref{fig:simulations} we plot the ratio ${f^\ast}/{s^\ast}$ against the number of unique patterns observed.
In the quartet and quintet case, the speedup is always greater than or equal to 1. 
This is guaranteed as, in those cases, the tree is always a caterpillar and the rank of the PLTs never becomes greater than 1.
In the sextet case, the two possible tree topologies are clearly noticeable (the caterpillar systematically giving a greater speedup than the balanced case).
For larger trees, the effect of different topologies on the cost is not noticeable in the plots as it is overpowered by sampling error. 
(In unpublished tests we could detect this difference by fixing sequence length and running many more trials.)
One may worry that \texttt{retroML} can sometimes take longer than $\mathcal{F}$, but it should be noted that this is only the case for short sequences, and in these cases the time taken to compute the likelihood is so small that this will not be of importance in practice.
It is the performance of \texttt{retroML} for large sequence lengths, where the overall computing time is longer, that shows its power.

The plots show clearly that \texttt{retroML} performs very well for small trees of up to 9 leaves.
However, for larger trees, the performance of \texttt{retroML} begins to degrade significantly.
It seems that the effectiveness of our method is confounded by the additional costs involved with high rank PLTs required by large trees.
Using the heuristics presented in Appendix~\ref{pseudo}, balanced trees present the worst case in this regard, with the maximum rank required bounded from above by $\log_2(m/2)$.
Although this is quite a good bound, we see from the plots that the effect of high rank PLTs is rather debilitating, and there seems little use in implementing \texttt{retroML} on trees with more than 15 leaves (this message is bourne out by unpublished results).
One would hope that for arbitrarily large trees there is some way of taking advantage of the favourable performance of \texttt{retroML} on small trees.
In \S\ref{sec:largetrees} we will discuss two approaches to how this may be achieved.

\subsection{Caterpillar trees, exact results}
For a caterpillar tree, since all the internal vertices are next to a leaf, the rank of the PLTs never becomes greater than one.
In this case the cost for a single pattern of our approach is exactly equal to that of standard approach.
This means that our approach will \emph{always} be faster when there is more than one site in the alignment.
In the quartet and quintet cases there is only the caterpillar topology, hence our approach will always be superior on trees of this size.
If every possible pattern is present in the alignment, we can calculate exactly the cost of \texttt{retroML} on a caterpillar tree, as follows.

The multiplicative cost $s^\ast_\sigma$ of \texttt{retroML} for each pattern with lookback $1\leq \sigma \leq (m\!-\!2)$ is $s_{\sigma}^\ast\!=\!\left[(m\!-\!\sigma)k\!+\! (m\!-\!2\!-\!\sigma)k^2\right]$, for $\sigma=0$ the cost is $s_0^\ast\!=\!\left[(m\!-\!1)k\!+\! (m\!-\!3)k^2\right]$, and for $\sigma=(m\!-\!1)$ the cost is $s_{m\!-\!1}^\ast\!=\!k$.
If every possible pattern occurs in the alignment, the total cost for any $m$ is then
\beqn
& s^\ast(\texttt{alldatacaterpillar})=\\ 
&\hspace{.5em} k\cdot[(m\!-\!1)k+(m\!-\!3)k^2]+\left(\sum_{\sigma =1}^{m-2} k^a(k\!-\!1)\cdot\left[(m\!-\!\sigma)k+(m\!-\!2\!-\!a)k^2 \right]\right)+k^{m\!-\!1}(k\!-\!1)\cdot k,\nonumber
\eqn
which, with a little finessing using geometric series, becomes
\beqn
s^\ast(\texttt{alldatacaterpillar})=k^3(k\!+\! 1)\sum_{i=0}^{m-3}k^i=k^3(k\!+\! 1)\frac{k^{m\!-\!2}\!-\!1}{k\!-\!1}.\nonumber
\eqn
Thus, in the case of a caterpillar tree and all possible patterns being present, the speedup that \texttt{retroML} provides is
\beqn
\frac{f^\ast(\texttt{alldatacaterpillar})}{s^\ast(\texttt{alldatacaterpillar})}=\frac{(k\!-\!1)k^{m\!-\!2}[(m\!-\!1)+(m\!-\!3)k]}{(k\!+\! 1)(k^{m\!-\!2}\!-\!1)}.\nonumber
\eqn
This is an $\mathcal{O}(mk)$ speedup.
For quartets $(m\!=\!4)$ this speedup works out to 4.48, exactly as indicated in Figure~\ref{fig:simulations}(a).
For quintets $(m\!=\!5)$ this speedup works out to 7.31, and for $m\!=\!50$ the speedup would be 140.

Of course, there is simply no way, even for modest values of $m$, that in practice a sequence alignment will contain any more than a (very) small number of the possible patterns.
For instance, taking $m\!=\!10$ and $k\!=\!4$, the number of possible patterns is $\sim\! 10^6$ and for $m\!=\!15$ the figure becomes $\sim\! 10^9$.
Even with maximally heterogeneous data, these numbers are well out of the reach of practical sequence alignments. 

%\subsection{Performance on ``bigfun''}
%
%Here we tabulate some results on the plant data set with trees chosen using the Yule model and subsets of sequences chosen randomly.
%
%\begin{itemize}
%\item $m=15$: Mean speedup was $1.65$ (4518 samples).
%\item $m=14$: Mean speedup was $1.78$ (100 samples).
%\item $m=13$: Mean speedup was $1.98$ (100 samples).
%\item $m=12$: Mean speedup was $2.09$ (100 samples).
%\item $m=11$: Mean speedup was $2.44$ (100 samples).
%\item $m=10$: Mean speedup was $2.79$ (100 samples).
%\item $m=9$: Mean speedup was $2.89$ (100 samples).
%\item $m=8$: Mean speedup was $3.27$ (100 samples).
%\item $m=7$: Mean speedup was $3.75$ (100 samples).
%\item $m=6$: Mean speedup was $4.15$ (100 samples).
%\item $m=5$: Mean speedup was $4.73$ (500 samples).
%\item $m=4$: Mean speedup was $3.97$ (100 samples).
%\end{itemize}

\section{Large trees}\label{sec:largetrees}

As we saw in \S\ref{sec:simulations}, \texttt{retroML} is only effective on realistic alignments when the number of sequences is less than about 16.
This is partly because the proportion of patterns present in an alignment of fixed length quickly becomes extremely small for large trees, but mostly because of the issues associated with the high rank PLTs required.
Clearly \texttt{retroML} will be very effective for quartet puzzling or other supertree methods reliant on performing maximum likelihood only on small subtrees.
We can only but recommend that in these cases our approach be adopted.

Here we will discuss how the favourable performance of our algorithm on small trees can be exploited to achieve speedups for maximum likelihood computations on arbitrarily large trees.
We have two different ideas as to how this could be achieved.

\subsubsection*{Divide}

The idea behind this approach is to divide a ``large'' tree of size $m$ into $q$ ``small'' subtrees of size close to $m/q$. 
For the root of each of these subtrees, we can use \texttt{retroML} to compute the PLV for each subpattern in the data set corresponding to that subtree.
Then, the likelihood of the whole tree can be computed in the usual way using Felsenstein's recursion $\mathcal{F}$ on the internal part of the tree.
The effectiveness of this approach hinges on the fact that, as the tree gets larger, the size of the internal part of the tree, that is to be computed using $\mathcal{F}$, increases at the same rate as the number of subtrees on the outer part that are to be computed using \texttt{retroML} (both effects are linear in $q$).
Thus the speedup obtained using our approach on the subtrees scales with the size of the whole tree and this will result in a tangible overall speedup for arbitrarily large trees.
%A schema for implementing this approach is represented in Figure~BLAH.

When implementing this approach, the PLVs at the root of each subtree and for each pattern must be retained in memory, and there will be additional bookkeeping involved in bringing together the relevant PLVs when it comes time to traverse the internal part of the tree.

Considering possible problems on the memory side of things, we note that in the standard implementation, $N$ likelihoods (at the root) and $2(m\!-\!1)k^2$ entries in the transition matrices must be retained in memory, which is an $\mathcal{O}(N\!+\! mk^2)$ memory requirement.
Employing the approach described above, using $q$ subtrees, requires us to record $(N_1\!+\! N_2\!+\! \ldots \!+\! N_q)$ PLVs, where $N_i$ is the number of unique subpatterns associated with the $i^\text{th}$ subtree.
In the (rather unlikely) worst case we have $(N_1\!+\! N_2\!+\! \ldots \!+\! N_q)=qN$, so the memory requirements on the internal part of the tree are $\mathcal{O}(qNk)$.
This could certainly be a large increase in memory requirements, but this is a very conservative estimate.
Additionally, it is linear in the size of the tree $m$ and hence will not get of hand for large trees.

With regard to numerical computations, for the standard implementation let $f_{\hat{m}}^\ast$ be the multiplicative cost of computing the PLV at the bifurcating root of a subtree of size $\hat{m}$.
We use our previous result that this cost is independent of tree topology and note that a caterpillar tree with a bifurcating root has $(m\!-\!2)$ half-internal vertices and a single cherry, so that
\beqn
f_{\hat{m}}^\ast\!=\! (\hat{m}\!-\!2)k(k\!+\! 1)+k.\nonumber
\eqn 
If, for subtrees of size $\hat{m}$, \texttt{retroML} provides an average speedup of $\texttt{su}_{\hat{m}}\!:=\!f_{\hat{m}}^\ast/s_{\hat{m}}^\ast$, then the speedup (on average) for a tree of size $m$ will be 
\beqn
\texttt{su}_{m} = \frac{N(qf_{\hat{m}}^\ast+(q\!-\!1)k(2k\!+\! 1)+k(3k\!+\! 1))}{N(\texttt{su}_{\hat{m}}^{-1}qf_{\hat{m}}^\ast+(q\!-\!1)k(2k\!+\! 1)+k(3k\!+\! 1))}.\nonumber
\eqn 
Taking the limit $q\rightarrow \infty$, we find that the speedup approaches
\beqn
\texttt{su}_m=\frac{\hat{m}\lambda}{(\hat{m}\lambda - 1)/\texttt{su}_{\hat{m}}+1},\nonumber
\eqn
where $\lambda \!:=\! (k\!+\! 1)/(2k\!+\! 1)$.
For instance, if we set $k\!=\!4$, $\hat{m}\!=\!6$ and assume a \texttt{retroML} speedup of $\texttt{su}_6=1,2,3,4$ and 5 (see Figure~\ref{fig:simulations}(c) for justification), the corresponding speedups achieved for extremely large trees would be 1, 1.83, 2.53, 3.13, 3.69 and 4.41, respectively. 

\subsubsection*{Restrict}\label{sec:restrict}

The idea behind this approach is to restrict the rank of the PLTs to some maximum value $R$.
This can be achieved by ``breaking'' the algorithm at any point where the next move will result in a PLT of rank greater than $R$, as follows

Suppose the method breaks at vertex $v$.
To continue, the PLT at $v$ is retained and the algorithm jumps directly to the next nearest leaf $\ell$.
Next, the PLT conditioned upon only the leaf state at $\ell$ is computed, and the standard procedure of computing PLTs is continued from here until vertex $v$ is reached.
At this point, the method combines (by multiplication on the relevant index) the current PLT with the PLT that was retained previously at $v$.
From here the method can continue as normal until the maximum rank is exceeded once again or the calculation terminates.
Computing likelihoods in this way can be made to give completely equivalent results to Felsenstein's approach.

\begin{figure}[tbp]
\centering

\includegraphics[scale=0.5]{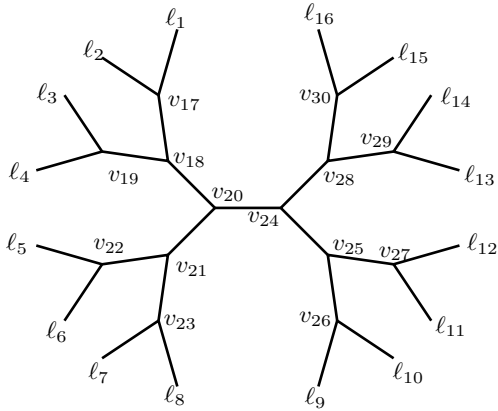}
\caption{Tree with 16 leaves}
\label{fig:16leaftree}
\end{figure}

We illustrate this idea by considering the multiplicative cost of computing the likelihood for a completely balanced tree (Figure~\ref{fig:16leaftree}) with $m\!=\!16$ leaves.
We do this using \texttt{retroML} with no restriction of the rank, which requires rank $r\!=\!3$ PLTs, using \texttt{retroML}${}_2$, which restricts to rank 2 PLTs, and (iii) using \texttt{retroML}${}_1$, which restricts to rank 1 PLTs.
The formulas for the PLTs in each case, alongside multiplicative costs, are presented in Appendix~\ref{retroMLrestrict}.
The attraction of (iii) is that this will result in a method that will \emph{always} be as fast or faster than Felsenstein's approach, no matter how little data there is.
This is because the additional computations required by high rank PLTs are completely avoided, whilst still taking advantage of the lookback structure of the patterns.
%We have labelled the internal vertices of the in the order that \texttt{retroML} traverses the tree. 

If every possible pattern occurs in the alignment, then, using the multiplicative costs presented in Appendix~\ref{retroMLrestrict} and setting $k\!=\!4$, we find that the speedup for case (i) is 35.78, for case (ii) the speedup is 35.84, and for case (iii) the speedup is 34.63.
This is a very positive result not least because restricting to rank $r\leq 1$ captures 97\% of the speedup of the best case (ii).
This is important for practical implementation because restricting to rank $r\leq 1$ will \emph{always} be as least as fast as $\mathcal{F}$, regardless of the size of the data set.

\begin{figure}[tbp]
\centering
\includegraphics[scale=.5]{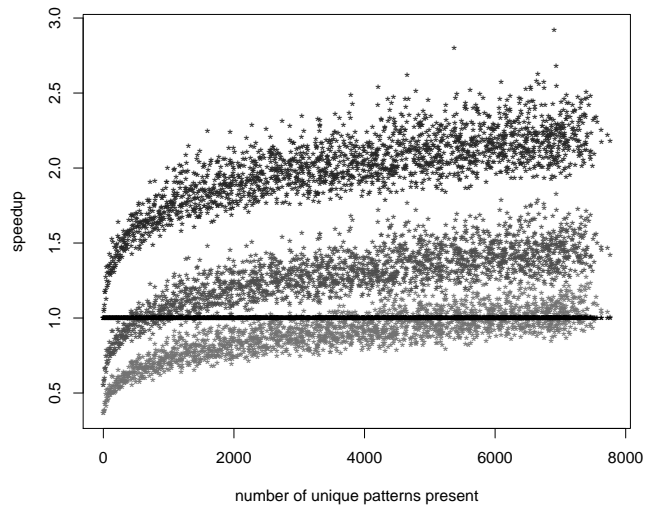} 
\caption{Speedups for balanced tree with $m\!=\!16$ leaves (2,001 trials, sequence length evenly spaced from 1 to $10^4$). Light grey is (unrestricted) $\texttt{retroML}$, dark grey is $\texttt{retroML}_2$, and black is $\texttt{retroML}_1$.}
\label{fig:restrict}
\end{figure}

To illustrate the effectiveness of this method for realistic circumstance, we tested the cases (i), (ii) and (iii) on simulated data using the tree in Figure~\ref{fig:16leaftree} and the model conditions described in \S\ref{sec:simulations}.
We found (Figure~\ref{fig:restrict}) that for realistic sequence lengths (1 to $10^4$), it is case (iii) that clearly performs best. 
This is because, for this range of sequence length, there is simply not enough data for the additional overhead of the unrestricted and rank 2 approaches to be worthwhile.

\section{Discussion}
%
%We say what we did.
In this article we have presented an alternative method for computing likelihoods in molecular phylogenetics. 
We have shown that with the simultaneous use of partial likelihood tensors and a lexicographic ordering of sites, we can achieve significant computational speedups.
We did this by a careful examination of the number of numerical computations that arise in both our method and the standard approach based on Felsenstein's recursion.
This type of analysis gives a refined perspective on the efficiency of the algorithms used in phylogenetics.
We showed that it is useful to do more than simply state orders of complexity; for instance we uncovered that the computational cost of an application of Felsenstein's formula at a particular vertex depends on where that vertex lies on the tree.
This observation has significant repercussions for anyone interested in writing efficient maximum likelihood software, and it is not clear which existing packages take account of this, as frequently the source code is unavailable.

A clear path for further analysis of our approach would be to pinpoint exactly under what circumstance it is worthwhile to use high rank PLTs. 
Our final example on a 16 leaf tree was telling, as even for the case of all possible patterns present in the data set, the optimal maximum rank was 2, not 3 as occurs naturally while using \texttt{retroML}.
One can envisage an algorithm that examines all the relevant structure of a sequence alignment and proceeds in the calculation of the likelihood using the absolute minimum number of computations. 
Considering the ``restrict'' modification of \texttt{retroML} we presented in \S\ref{sec:restrict}, one could plausibly design a more refined method which, for each vertex on the tree, sets the maximum rank of the PLTs depending upon the density of data associated with the subtree subtended by that vertex.
This may be difficult to implement and of theoretical interest only, but it raises the question of exactly what is the minimum number of calculations required to compute the likelihood of a particular data set.
A clue to the impracticality of such a scheme comes from \citet{stamatakis2005a} and their decision to only consider homogeneous subpatterns; sometimes the cost of extra pre-computation required by a clever method is simply not worth all the trouble. 

Although not examined in this article, PLTs also hold significant promise for improving performance under tree perturbations.
This is because the partial likelihood tensors can encode a substantial amount of information that is invariant under tree perturbations.
For instance, consider the rank 2 PLT represented in Figure~\ref{fig:PLTs}(b).
Suppose that there was a larger subtree rooted at $v_{10}$, and we were interested in tree perturbations that altered this subtree and the leaves $\ell_5,\ell_6$ and $\ell_7$, but left the rest of the tree unchanged.
Clearly this PLT is invariant to any such perturbations and could be retained throughout.
In this way, the power of high rank PLTs could be applied directly to efficient calculations during tree perturbation in close analogy to the ``tree swapping'' technique of \citet{guindon2003} using PLVs.

\subsubsection*{Acknowledgements}
We thank Ran Libeskind-Hadas for helpful discussions and Michael Sumner for reading an early draft.
This work was conducted with support from Australian Research Council grants DP0770991 and DP0877447.

%We say it is clever because.
%We say it matters because.
%We say how it is unfinished but gives a new perspective on the problem and by presenting exact results we have raised the bar in complexity theory of phylogenetic likelihood.
%Previously in other work orders of complexity have been given, we have shown clearly that the associated constants can make a real difference; something that a practicing phylogeneticist can get a real benefit from.
%In principle one could look at the data extensively and determine the appropriate way to use the PLTs.
%We gave two suggestions on this.
%
%Another avenue of future research is applying our approach to local searches in tree space.
%It is clear that clever use of the special properties of higher rank PLTs could be very effective in this problem.
%

\begin{appendix}
\setcounter{equation}{0}
\renewcommand{\theequation}{{A}-\arabic{equation}}
\def\faveChild{\texttt{favouriteChild}} 
\def\traversetree{\texttt{getTraversalOrder}} 
\def\badChild{\texttt{badChild}} 
\def\bestStartingLeaf{\texttt{bestStartingLeaf}} 
\def\retroML{\texttt{retroML}} 

\def\process{\texttt{process}} 
\def\processgc{\texttt{processGoodChild}} 
\def\processbc{\texttt{processBadChild}}

\section{Restricted \texttt{retroML}}\label{retroMLrestrict}

Here we present the computation of PLTs conditioned on subpatterns of size 1 to 16 for the phylogenetic tree Figure~\ref{fig:16leaftree} using \texttt{retroML} and its ``restricted'' versions \texttt{retroML}$_2$ and \texttt{retroML}$_1$.

(i) With no restriction upon the rank of the PLTs, \texttt{retroML} proceeds as displayed in Table~\ref{Fig:norestrict}.

(ii) With rank restricted to $r\leq 2$, \texttt{retroML}$_2$ proceeds by ``breaking'' in order to visit $\ell_5$ and to visit $\ell_9$ as displayed in Table~\ref{fig:restrict2}.

(iii) With rank restricted to $r\leq 1$, \texttt{retroML}$_1$ proceeds by ``breaking'' in order to visit the leaves $\ell_3,\ell_5,\ell_7,\ell_9,\ell_{11}$ and $\ell_{13}$ as displayed in Table~\ref{fig:restrict1}.

\begin{table}[tb]
\noindent\begin{tabular}{p{.8\textwidth}l}
$\Psi_{a}^{(v_1)}(X_1) = M_{aX_1}^{(1)} $ & $ s^\ast =0 $ \\
$\Psi_{ab}^{(v_{18},v_{19})}(X_1X_2) = \left( \sum_{a'} M_{aa'}^{(17)}\Psi_{a'}^{(v_1)}(X_1)M_{a'X_2}^{(2)} \right)M_{ab}^{(19)} $ & $ s^\ast =2k^2 +k $ \\
$\Psi_{ab}^{(v_{18},v_{19})}(X_1X_2X_3) = \Psi_{ab}^{(v_{18},v_{19})}(X_1X_2)M_{bX_3}^{(3)} $ & $ s^\ast = k^2 $ \\ 
$\Psi_{abc}^{(v_{20},v_{21},v_{22})}(X_1X_2X_3X_4)$ & $ s^\ast = k^3+3k^2 $ \\
\hspace{1em} $ = \left(\sum_{a'} M_{aa'}^{(18)}\left(\sum_{b'}\Psi_{a'b'}^{(v_{18},v_{19})}(X_1X_2X_3)M_{b'X_{4}}^{(4)}\right)\right)M_{ab}^{(21)} M_{bc}^{(22)} $ \\  
$\Psi_{abc}^{(v_{20},v_{21},v_{22})}(X_1\ldots X_5) = \Psi_{abc}^{(v_{20},v_{21},v_{22})}(X_1X_2X_3X_4) M_{cX_5}^{(5)} $ & $  s^\ast = k^3 $ \\
$\Psi_{ab}^{(v_{20},v_{23})}(X_1\ldots X_6) = \sum_{b'}\left(\sum_{c'} \Psi_{ab'c'}^{(v_{20},v_{21},v_{22})}(X_1\ldots X_5)M_{c'X_6}^{(6)}\right) M_{b'b}^{(23)} $ & $ s^\ast = 2k^3 $ \\
$\Psi_{ab}^{(v_{20},v_{23})}(X_1\ldots X_7) = \Psi_{ab}^{(v_{20},v_{23})}(X_1\ldots X_6)M_{bX_7}^{(7)} $ & $ s^\ast = k^2 $ \\
$\Psi_{abc}^{(v_{24},v_{25},v_{26})}(X_1\ldots X_8)$ & $ s^\ast = k^3+3k^2 $ \\
\hspace{1em} $= \left(\sum_{a'}M_{aa'}^{(20)}\left(\sum_{b'}\Psi_{a'b'}^{(v_{20},v_{23})}(X_1\ldots X_7)M_{b'X_8}^{(8)}\right)\right)M_{ab}^{(25)}M_{bc}^{(26)} $ \\
$\Psi_{abc}^{(v_{24},v_{25},v_{26})}(X_1\ldots X_9) = \Psi_{abc}^{(v_{24},v_{25},v_{26})}(X_1\ldots X_8)M_{cX_9}^{(9)} $ & $  s^\ast = k^3 $ \\
$\Psi_{ab}^{(v_{24},v_{27})}(X_1\ldots X_{10}) =   \sum_{b'}\left(\sum_{c'} \Psi_{ab'c'}^{(v_{24},v_{25},v_{26})}(X_1\ldots X_9)M_{c'X_{10}}^{(10)}\right) M_{b'b}^{(27)}  $ & $ s^\ast = 2k^3 $ \\
$\Psi_{ab}^{(v_{24},v_{27})}(X_1\ldots X_{11}) = \Psi_{ab}^{(v_{24},v_{27})}(X_1\ldots X_{10})M_{bX_{11}}^{(11)} $ & $ s^\ast = k^2 $ \\
$\Psi_{ab}^{(v_{28},v_{29})}(X_1\ldots X_{12}) = \left( \sum_{a'} M_{aa'}^{(24)}\left(\sum_{b'}\Psi_{a'b'}^{(v_{24},v_{27})}(X_1\ldots X_{11})M_{b'X_{12}} \right)\right)M_{ab}^{(29)} $ & $ s^\ast = 3k^2 $ \\
$\Psi_{ab}^{(v_{28},v_{29})}(X_1\ldots X_{13}) = \Psi_{ab}^{(v_{28},v_{29})}(X_1\ldots X_{12})M_{bX_{13}}^{(13)} $ & $ s^\ast = k^2 $ \\
$\Psi_{a}^{(v_{30})}(X_1\ldots X_{14}) = \sum_{a'} M_{aa'}^{(28)}\left(\sum_{b'}\Psi_{a'b'}^{(v_{28},v_{29})}(X_1\ldots X_{13})M_{b'X_{14}}^{(14)}\right) $ & $ s^\ast = 2k^2 $ \\
$\Psi_{a}^{(v_{30})}(X_1\ldots X_{15}) = \Psi_{a}^{(v_{30})}(X_1\ldots X_{14})M_{aX_{15}}^{(15)} $ & $ s^\ast = k $ \\
$\Psi_{a}^{(v_{30})}(X_1\ldots X_{16}) = \Psi_{a}^{(v_{30})}(X_1\ldots X_{15})M_{aX_{16}}^{(16)} $ & $ s^\ast = k $ \\
\end{tabular}
\caption{Calculating PLTs for balanced tree with no restriction on rank.}
\label{Fig:norestrict}
\end{table}

\begin{table}[tb]
\noindent
\begin{tabular}{p{.7\textwidth}l}
$\Psi_{a}^{(v_1)}(X_1) = M_{aX_1}^{(1)}$ & $s^\ast =0$ \\
$\Psi_{ab}^{(v_{18},v_{19})}(X_1X_2) = \left( \sum_{a'} M_{aa'}^{(17)}\Psi_{a'}^{(v_1)}(X_1)M_{a'X_2}^{(2)} \right)M_{ab}^{(19)}$ & $s^\ast =2k^2 +k$ \\
$\Psi_{ab}^{(v_{18},v_{19})}(X_1X_2X_3) = \Psi_{ab}^{(v_{18},v_{19})}(X_1X_2)M_{bX_3}^{(3)}$ & $s^\ast = k^2$ \\ 
$\Psi_{ab}^{(v_{20},v_{21})}(X_1X_2X_3X_4)$ & $s^\ast = 3k^2$ \\
\hspace{1em} $= \left(\sum_{a'} M_{aa'}^{(18)}\left(\sum_{b'}\Psi_{a'b'}^{(v_{18},v_{19})}(X_1X_2X_3)M_{b'X_{4}}^{(4)}\right)\right)M_{ab}^{(21)}$ \\
$\Psi_{a}^{(v_{22})}(X_5) = M_{aX_5}^{(5)}$ & $s^\ast = 0$ \\
$\Psi_{ab}^{(v_{20},v_{23})}(X_1\ldots X_6)$ & $s^\ast = k^3+2k^2+k$ \\ 
\hspace{1em} $= \sum_{b'} \Psi_{ab'}^{(v_{20},v_{21})}(X_1X_2X_3X_4)\left(\sum_{c'}M_{b'c'}^{(22)}\Psi_{c'}^{(v_{22})}(X_5)M_{c'X_6}^{(6)}\right)M_{b'b}^{(23)}$  \\
$\Psi_{ab}^{(v_{20},v_{23})}(X_1\ldots X_7)  = \Psi_{ab}^{(v_{20},v_{23})}(X_1\ldots X_6) M_{bX_7}^{(7)} $ & $s^\ast = k^2$ \\ 
$\Psi_{ab}^{(v_{24},v_{25})}(X_1\ldots X_8)$ & $s^\ast = 3k^2$  \\ 
\hspace{1em} $= \left(\sum_{a'}M_{aa'}^{(20)}\left(\sum_{b'}\Psi_{a'b'}^{(v_{20},v_{23})}(X_1\ldots X_7)M_{b'X_8}^{(8)}\right)\right) M_{ab}^{(25)}$ \\
$\Psi_{a}^{(v_{26})}(X_9) = M_{aX_{9}}^{(9)}$ & $s^\ast = 0$ \\
$\Psi_{ab}^{(v_{24},v_{27})}(X_1\ldots X_{10})$ & $s^\ast = k^3+2k^2+k$ \\
\hspace{1em} $ = \sum_{b'}\Psi_{ab'}^{(v_{24},v_{25})}(X_1\ldots X_8)\left(\sum_{c'}M_{b'c'}^{(26)}\Psi_{c'}^{(v_{26})}(X_9)M_{c'X_{10}}^{(10)}\right)M_{b'b}^{(27)}$  \\ 
$\Psi_{ab}^{(v_{24},v_{27})}(X_1\ldots X_{11}) = \Psi_{ab}^{(v_{24},v_{27})}(X_1\ldots X_{10})M_{bX_{11}}^{(11)}$ & $s^\ast = k^2$ \\
$\Psi_{ab}^{(v_{28},v_{29})}(X_1\ldots X_{12})$ & $s^\ast = 3k^2$ \\
\hspace{1em} $= \left(\sum_{a'}M_{aa'}^{(24)}\left(\sum_{b'}\Psi_{a'b'}^{(v_{24},v_{27})}(X_1\ldots X_{11})M_{b'X_{12}}^{(12)}\right)\right) M_{ab}^{(29)}$ \\
$\Psi_{ab}^{(v_{28},v_{29})}(X_1\ldots X_{13}) = \Psi_{ab}^{(v_{28},v_{29})}(X_1\ldots X_{12})M_{bX_{13}}^{(13)}$ & $s^\ast = k^2$ \\
$\Psi_{a}^{(v_{30})}(X_1\ldots X_{14}) = \sum_{a'} M_{aa'}^{(28)}\left(\sum_{b'}\Psi_{a'b'}^{(v_{28},v_{29})}(X_1\ldots X_{13})M_{b'X_{14}}^{(14)}\right)$ & $s^\ast = 2k^2$ \\
$\Psi_{a}^{(v_{30})}(X_1\ldots X_{15}) = \Psi_{a}^{(v_{30})}(X_1\ldots X_{14})M_{aX_{15}}^{(15)}$ & $s^\ast = k$ \\
$\Psi_{a}^{(v_{30})}(X_1\ldots X_{16}) = \Psi_{a}^{(v_{30})}(X_1\ldots X_{15})M_{aX_{16}}^{(16)}$ & $s^\ast = $k \\
\end{tabular}
\caption{Calculating PLTs for balanced tree while restricted to rank $r\leq 2$.}
\label{fig:restrict2}
\end{table}

\begin{table}[tb]
\noindent\begin{tabular}{p{.7\textwidth}l}
$\Psi_{a}^{(v_1)}(X_1) = M_{aX_1}^{(1)}$ & $s^\ast =0$ \\
$\Psi_{a}^{(v_{18})}(X_1X_2) = \sum_{a'} M_{aa'}^{(17)}\Psi_{a'}^{(v_1)}(X_1)M_{a'X_2}^{(2)}$  & $s^\ast =k^2 +k$ \\
$\Psi_{a}^{(v_{19})}(X_3) = M_{bX_3}^{(3)}$ & $s^\ast = 0$ \\ 
$\Psi_{a}^{(v_{20})}(X_1X_2X_3X_4)$ & $s^\ast = 2k^2+2k$\\
\hspace{1em}  $= \sum_{b'}M_{ab'}^{(18)}\Psi_{b'}^{(v_{18})}(X_1X_2) \left(\sum_{a'} M_{b'a'}^{(19)}\Psi_{a'}^{(v_{19})}(X_3)M_{a'X_{4}}^{(4)}\right)$  \\
$\Psi_{a}^{(v_{22})}(X_5) = M_{aX_5}^{(5)}$ & $s^\ast = 0$ \\
$\Psi_{a}^{(v_{21})}(X_5X_6) = \sum_{a'}M_{aa'}^{(22)}\Psi_{a'}^{(v_{22})}(X_5)M_{a'X_6}^{(6)}$ & $s^\ast = k^2+k$ \\
$\Psi_{a}^{(v_{23})}(X_7) = M_{aX_7}^{(7)}$ & $s^\ast = 0$ \\
$\Psi_{a}^{(v_{24})}(X_1\ldots X_8)$ & $s^\ast = 3k^2+3k$ \\
\hspace{1em} $= \sum_{c'}M_{ac'}^{(20)}\Psi_{c'}^{(v_{20})}(X_1X_2X_3X_4)\left(\sum_{b'} M_{c'b'}^{(21)}\Psi_{b'}^{(v_{21})}(X_5X_6)\left(\sum_{a'}M_{b'a'}^{(23)}\Psi_{a'}^{(v_{23})}(X_7)M_{a'X_8}\right)\right)$    \\
$\Psi_{a}^{(v_{26})}(X_9) = M_{aX_{9}}^{(9)}$ & $s^\ast = 0$ \\
$\Psi_{a}^{(v_{25})}(X_9X_{10}) = \sum_{a'}M_{aa'}^{(26)}\Psi_{a'}^{(v_{26})}(X_9)M_{a'X_{10}}^{(10)}$ & $s^\ast = k^2+k$ \\
$\Psi_{a}^{(v_{27})}(X_{11}) = M_{aX_{11}}^{(11)}$ & $s^\ast = 0$ \\
$\Psi_{a}^{(v_{28})}(X_1\ldots X_{12})$ & $s^\ast = 3k^2+3k$ \\
\hspace{1em} $ = \sum_{c'}M_{ac'}^{(28)}\Psi_{c'}^{(v_{24})}(X_1\ldots X_8)\left(\sum_{b'}M_{c'b'}^{(25)}\Psi_{b'}^{(v_{25})}(X_9X_{10})\left(\sum_{a'}M_{b'a'}^{(27)}\Psi_{a'}^{(v_{27})}(X_{12})M_{a'X_{12}}^{(12)}\right)\right)$  \\
$\Psi_{a}^{(v_{29})}( X_{13}) = M_{aX_{13}}^{(13)}$ & $s^\ast = 0$ \\
$\Psi_{a}^{(v_{30})}(X_1\ldots X_{14})$ & $s^\ast = 2k^2+2k$ \\
\hspace{1em}$= \sum_{b'}M_{ab'}^{(28)}\Psi_{b'}^{(v_{28})}(X_1\ldots X_{12})\left(\sum_{a'}M_{b'a'}^{(29)}\Psi_{a'}^{(v_{29})}( X_{13})M_{a'X_{14}}^{(14)}\right) $ \\
$\Psi_{a}^{(v_{30})}(X_1\ldots X_{15}) = \Psi_{a}^{(v_{30})}(X_1\ldots X_{14})M_{aX_{15}}^{(15)}$ & $s^\ast = k$ \\
$\Psi_{a}^{(v_{30})}(X_1\ldots X_{16}) = \Psi_{a}^{(v_{30})}(X_1\ldots X_{15})M_{aX_{16}}^{(16)}$ & $s^\ast = k$ \\
\end{tabular}
\caption{Calculating PLTs for balanced tree while restricted to rank $r\leq 1$.}
\label{fig:restrict1}
\end{table}

\section{Pseudo-code}\label{pseudo}

Here we present pseudo-code for implementing \texttt{retroML} generically.
First we present heuristics for finding the best way to traverse the tree given that it is best to minimize the rank of PLTs that arise during implementation.

Given a vertex $v$, we define $h_{\mbox{min}}(v)$ as the minimum distance to a leaf:
\beqn
h_{\mbox{min}}(v):=\min_{\ell\in L} (|P(v,\ell)|),\nonumber
\eqn
where $L$ is the leaf vertices and $P(v,u)$ is the tuple of vertices that lie on the path from vertex $v$ to $u$.
We denote the subtree subtended by a vertex $v$ as $T_v$.

The $\faveChild$ of a vertex is chosen first based on minimum $h_{\mbox{min}}$; next, if this is not unique, based on the smallest subtended subtree; finally, on minimizing the sum of all the $h_{\mbox{min}}$ on all vertices in the subtended subtree.
Given a starting leaf, $\traversetree$ returns the traversal of a tree that \texttt{retroML} takes by following the $\faveChild$ heuristic. 

Another heuristic, $\bestStartingLeaf$, attempts to find the optimal place for the \texttt{retroML} method to start.
This is done by taking into consideration the comments of \S\ref{sec:lookbacks}.
We choose leaves such that the maximum $h_{\mbox{min}}$ of the subtrees hanging off the path $P(\ell_0,\ell_\dag)$ is minimized, and then we ensure that the maximum $h_{\mbox{min}}$ of these subtrees occurs early in the tree traversal.

\texttt{retroML} itself proceeds given a tree $T$ and a transition matrix $M_{ab}^{(v)}$ for every vertex excluding the root.
For optimal performance, the patterns in the alignment should be sorted lexicographically with respect to the order of leaves in \traversetree. 

\parbox[h]{80em}{
\begin{tabbing} 
~~~\=~~~\=~~~\=~~~\=~~~\=~~~\=\kill 
\faveChild($v$) \\ 
\> \textbf{let} the children of $v$ be $u_{1},u_2$ \\ 
\> \textbf{if} $h_{\mbox{min}}(u_{i})$ has a unique minimum at $u_{a}$ \\ 
\> \> \textbf{return} $u_{a}$ \\ 
\> \textbf{else} if $|T_{u_i}|$ has a unique minimum at $u_{b}$ \\ 
\> \> \textbf{return} $u_{b}$ \\ 
\> \textbf{else} \textbf{if} $\left(\sum_{v \in T_{u_i}}h_{\mbox{min}}(v)\right)$ 
has a unique minimum at $u_{c}$ \\ 
\> \> \textbf{return} $u_{c}$ \\ 
\> \textbf{else} \\ 
\> \> \textbf{return} arbitrarily chosen $u_{i}$ \\ 
\end{tabbing}  
}

\parbox[h]{80em}{
\begin{tabbing} 
~~~\=~~~\=~~~\=~~~\=~~~\=~~~\=\kill 
\badChild($v$) \\ 
\> \textbf{let} the children of $v$ be $u_{1},u_2$ \\ 
\> \textbf{if} $u_1=\faveChild(v)$ \\
\> \> \textbf{return} $u_2$ \\
\> \textbf{else} \\
\> \> \textbf{return} $u_1$ 
\end{tabbing}
}

\parbox[h]{80em}{
\begin{tabbing} 
~~~\=~~~\=~~~\=~~~\=~~~\=~~~\=\kill 
\traversetree($\ell_{0}$) \\ 
\> mark $\ell_0$ as ``visited'' and all other vertices as ``unvisited'' \\ 
\> \textbf{let} $v$ be the current vertex, so $v\leftarrow\ell_{0}$ \\ 
\> \textbf{let} $\xi$ be an ordering of vertices in $T$; initially $\xi\leftarrow (\ell_0)$ \\ 
\> \> [this will eventually be the traversal order, 
$(\ell_{0},\ldots,\ell_{\dag})$] \\ 
\> \textbf{while} there are any unvisited vertices \\ 
\> \> \textbf{if} $v$ has any children \\ 
\> \> \> $v\leftarrow$ \faveChild($v$) \\ 
\> \> \textbf{else} \\ 
\> \> \> set $v$ to be the closest unvisited vertex to $v$ \\ 
\> \> mark $v$ as visited and append it to $\xi$ \\ 
\> \textbf{return} $\xi$ 
\end{tabbing} 
}

\parbox[h]{80em}{
\begin{tabbing}   
~~~\=~~~\=~~~\=~~~\=~~~\=~~~\=\kill   
\texttt{\bestStartingLeaf} \\   
\> \textbf{let} $L$ be a set of leaves, each chosen arbitrarily from a cherry in $T$ \\   
\> \textbf{for} each $\ell_{i} \in L$ \\   
\> \> compute $(\ell_{i},\ldots,\ell_{i\dag})=$ \traversetree$(\ell_i)$ \\  
\> \> \textbf{let} $P^{(i)}$ be the tuple $(v_{1}\ldots v_{k(i)})$ of vertices $v_{j}$ that lie on the path from $\ell_i$ to $\ell_{i\dag}$ \\  
\> \> \textbf{let} $S^{(i)}$ be a tuple $(s_{1},\ldots , s_{k(i)})$ of min heights: $s_j\leftarrow h_{\mbox{min}}(v_j)$ \\  
\> \> \textbf{let} $q_{i} = \max(s_{j})$ \\   
\> \textbf{let} $q_0=\min q_i$ \\  
\> remove all leaves $\ell_{i}$ from $L$ such that $ q_i > q_0 $ \\  
\> \textbf{if} $L=\{\ell \}$ return $\ell$ \\  
\> \textbf{for} each remaining $\ell_i$ in $L$ \\   
\> \> consider each tuple $S^{(i)}=(s_1,s_2,\ldots, s_{k(i)})$ as an integer: $b_i\leftarrow s_1s_2\ldots s_{k(i)}$ \\  
\> \textbf{return} $\ell_{i}$ for which $b_{i}$ is maximal, ties resolved arbitrarily. 
\end{tabbing}   
}

\parbox[h]{80em}{
\begin{tabbing} 
~~~\=~~~\=~~~\=~~~\=~~~\=~~~\=\kill 
\retroML \\
\> \textbf{let} $L$ be leaves of tree $T$\\
\> compute $\ell_0\leftarrow$\bestStartingLeaf$(T)$ \\
\> reroot $T$ at $\ell_0$\\
\> \textbf{let} $\Psi_a \leftarrow M^{(\ell_0)}_{aX_{l_0}}$; this is the current working PLT\\
\> \textbf{let} $\Omega$ be a list; this will be the list ``lookback'' PLTs \\
\> \textbf{let} $\Omega_1 \leftarrow \Psi_a$ \\
\> \textbf{let} $v\leftarrow \faveChild(\ell_0)$; this is the current working vertex \\
\> \> [now we get the method started by processing the first pattern $\texttt{patt}_1$] \\
\> $\texttt{process}(v,\texttt{patt}_1)$ \\
\> \textbf{while} there are patterns left to process \\
\> \> get lookback $\sigma$ of current pattern $\texttt{patt}_i$ \\
\> \> \textbf{let} $\Psi \leftarrow \Omega_{m-\sigma}$ \\
\> \> suppose $\Psi=\Psi^{(v_1,v_2,\ldots,v_r)}_{a_1a_2\ldots a_r}$ \\
\> \> \textbf{let} $v\leftarrow v_1 $ \\
\> \> $\texttt{process}(v,\texttt{patt}_i)$ \\
\end{tabbing}
}

\parbox[h]{80em}{
\begin{tabbing} 
~~~\=~~~\=~~~\=~~~\=~~~\=~~~\=\kill 
\process$(v,\texttt{patt})$ \\
\> \textbf{if} $v$ is not a leaf \\
\> \> \processgc$(v,\texttt{patt})$ \\
\> \> \processbc$(v,\texttt{patt})$ \\
\end{tabbing}
}

\parbox[h]{80em}{
\begin{tabbing} 
~~~\=~~~\=~~~\=~~~\=~~~\=~~~\=\kill 
\processgc$(v,\texttt{patt})$ \\
\> $v\leftarrow \faveChild(v)$ \\
\> \textbf{if} $v$ is a leaf \\ 
\> \> append $\Psi$ to $\Omega$ \\
\> \> $\Psi_{a_1\ldots a_r}^{(v_1,\ldots,v_r)}\leftarrow \Psi_{a_1\ldots a_r}^{(v_1,\ldots,v_r)}M_{a_1X_v}^{(v)}$; [see Figure~\ref{fig:moves}(a)] \\
\> \textbf{else} \\
\> \> $\Psi_{aa_2\ldots a_r}^{(v,v_2,\ldots,v_r)}\leftarrow \Psi_{a_2\ldots a_r}^{(v_2,\ldots,v_r)}M_{a_2a}^{(v)}$; [see Figure~\ref{fig:moves}(c)]  \\
\> \> \process$(v,\texttt{patt})$ \\
\end{tabbing}
}

\parbox[h]{80em}{
\begin{tabbing} 
~~~\=~~~\=~~~\=~~~\=~~~\=~~~\=\kill 
\processbc$(v,\texttt{patt})$ \\
\> $v\leftarrow \texttt{badChild}(v)$ \\
\> \textbf{if} $v$ is a leaf \\ 
\> \> append $\Psi$ to $\Omega$ \\
\> \> $\Psi_{a_1\ldots a_r}^{(v_1,\ldots,v_r)}\leftarrow \sum_a\Psi_{aa_1\ldots a_{r}}^{(u,v_1,\ldots,v_r)}M_{aX_v}^{(v)}$; [see Figure~\ref{fig:moves}(b)]  \\
\> \textbf{else} \\
\> \> $\Psi_{a_1a_2\ldots a_r}^{(v,v_2,\ldots,v_r)}\leftarrow \sum_a\Psi_{aa_2\ldots a_{r}}^{(u,v_2,\ldots,v_r)}M_{aa_1}^{(v)}$; [see Figure~\ref{fig:moves}(d)]  \\
\> \> \process$(v,\texttt{patt})$ \\
\end{tabbing}
}

\end{appendix}

\bibliographystyle{jtbnew}
\bibliography{C:/reference/masterAB}

\end{document}